\documentclass[useAMS,usenatbib]{mn2e}

\usepackage{natbib}
\usepackage{cite}
\usepackage{amssymb}
\usepackage{graphicx}
\usepackage{epsf}
\usepackage{longtable}[1]
\usepackage{graphics}
\usepackage{lscape}
\usepackage{color}
\newcommand{\mnras}{MNRAS}
\newcommand{\apj}{ApJ}
\newcommand{\apjl}{ApJL}
\newcommand{\apjs}{ApJS}
\newcommand{\aj}{AJ}
\newcommand{\aap}{A\&A}
\newcommand{\araa}{ARAA}
\newcommand{\pasp}{PASP}
\newcommand{\aaps}{AAPS}

\newcommand{\apss}{Ap\&SS}
\newcommand{\aspc}{ASPC}

\newcommand{\ha}{H$\alpha$ \ }
\newcommand{\sex}{\emph{SExtractor}\ }
\newcommand{\mstar}{M\ensuremath{_{\star}}}
\newcommand{\msol}{M\ensuremath{_{\small{\sun}}}}
\newcommand{\msun}{M\ensuremath{_{\small{\sun}}}}
\newcommand{\zsol}{Z$_{\small{\sun}}$}

\newcommand{\msunyr}{M$_\odot$yr$^{-1}$}

\title[On the Fraction of Star Clusters Surviving the Embedded Phase]{On the Fraction of Star Clusters Surviving the Embedded Phase}
\author[Q. E. Goddard, N. Bastian, and R. C. Kennicutt]{Q. E. Goddard$^{1}$\thanks{E-mail:
goddard@ast.cam.ac.uk; qeg20@cam.ac.uk},  N. Bastian$^{1}$ \& R. C. Kennicutt$^{1}$\\
$^{1}$Institute of Astronomy, University of Cambridge, Madingley Road, Cambridge. CB3 0HA}

\begin{document}

\date{}

\pagerange{\pageref{firstpage}--\pageref{lastpage}} \pubyear{2009}

\maketitle

\label{firstpage}

\begin{abstract}

In this paper we derive ages and masses for 276 clusters in the merger galaxy NGC 3256. This was achieved by taking accurate photometry in four wavebands from archival HST images. Photometric measurements are compared to synthetic stellar population (SSP) models to find the most probable age, mass and extinction. The cluster population of NGC 3256 reveals an increase in the star formation rate over the last 100 million years and the initial cluster mass function (ICMF) is best described by a power law relation with slope $\alpha = 1.85 \pm 0.12$.

Using the observed cluster population for NGC 3256 we calculate the implied mass of clusters younger than 10 million years old, and convert this to a cluster formation rate over the last 10 million years. Comparison of this value with the star formation rate (SFR) indicates the fraction of stars found within bound clusters after the embedded phase of cluster formation, $\Gamma$, is $22.9\% \pm^{7.3}_{9.8} $ for NGC 3256. We carried out an in-depth analysis into the errors associated with such calculations showing that errors introduced by the SSP fitting must be taken into account and an unconstrained metallicity adds to these uncertainties. Observational biases should also be considered.

Using published cluster population data sets we calculate $\Gamma$ for six other galaxies and examine how $\Gamma$ varies with environment. We show that $\Gamma$ increases with the star formation rate density and can be described as a power law type relation of the form $\Gamma(\%) = (29.0\pm{6.0}) \Sigma_{SFR}^{0.24\pm0.04} (\msol \ yr^{-1}\ kpc^{-2})$.


\end{abstract}


\section{Introduction}

Galaxy mergers produce some of the most extreme star formation events in the universe, inducing starbursts with intensities (i.e. star-formation rates; SFRs) orders of magnitude larger than in quiescent galaxies.  In the distant universe these can be seen as Hyper-luminous Infrared galaxies \citep[e.g.][]{verma02} with SFRs exceeding a few thousand solar masses per year.  In the local universe, ongoing mergers like Arp~220 e.g. \citep[e.g.][]{scoville98,wilson06} and NGC~3256 \citep[e.g.][]{zepf99,trancho07b} have SFRs of a few tens to hundreds of solar masses per year, and due to their proximity, can be resolved with Hubble Space Telescope imaging. In these systems large numbers of super-star clusters are often found, with ages of a 1-500~Myr and masses up to $\sim 8\times10^7$\msol\ \citep{maraston04}.  We can think of star clusters as simple stellar populations which can be modelled relatively easily. This, combined with their high surface brightness that allows them to be easily  detected, and their long lifetimes, make them ideal tracers of the star formation history of a galaxy.

NGC 3256 is a relatively nearby starburst galaxy that is clearly the result of a recent galactic merger. It displays two prominent tidal tails, thought to have been produced during the first encounter between two spiral galaxies approximately 500 million years ago \citep{zepf99,english03}. \citet{zepf99} catalogued over 1000 young star clusters in the main body of the merger using HST imaging.  \citet{trancho07b} discovered three massive ($1-3 \times 10^5$ \msol) clusters within one of the tidal tails, whose ages and velocities places their formation within the tidal debris. Additionally, \citet{trancho07a} (hereafter T07a) studied a further sample of 23 clusters in the main body of the remnant and found that the clusters had metallicities of $\sim1.5$ \zsol, masses in the range $2-4 \times 10^5 $\msol\ and ages from a few to 150 Myr.  The current SFR of NGC~3256 is $\sim50$~\msunyr \citep{bastian08}, however it's SFR appears to have been increasing for the past $\sim200$~Myr (see \S~\ref{sec:formation-history}) and \citet{trancho07a} argue that it is likely to continue to increase in the future.


The age distribution of clusters reveals the underlying star formation history of the host galaxy. The mass distribution of clusters gives information on the cluster initial mass function (CIMF) which is often approximated by a power-law of the form $Ndm \propto M^{-\alpha}dm$, with $\alpha \sim 2$ \citep{elmegreen97,zhang99,degrijs03b,bik03}. More recently, it has been shown that the CIMF may have a truncation at high masses, being well described by a Schechter function \citep{schechter76} which is a power-law in the low-mass regime and has an exponential cutoff above a given mass, \mstar \citep{gieles06a,bastian08,larsen09,gieles09}. 

In order to obtain accurate cluster ages and masses either spectroscopy or photometry may be used. Although spectra yield more accurate results, photometry is more applicable to large sample sets. In order to break the age-dust degeneracy, a cluster must be observed in at least four broad photometric bands and cover the Balmer break \citep{anders04}.   By comparing the observed colours and magnitudes of each cluster to synthetic stellar population (SSP) models, we can estimate the age, mass and extinction of each cluster. This technique has been used on several cluster populations, the Antennae \citep{fall05, anders07}, NGC 1569 \citep{anders04} \& M51 \citep{bastian05b}. 

If all stars are formed in clusters, then only a small percentage of them are still within clusters at the end of the embedded phase \citep{lada03}.  For up to the first three million years of a cluster's lifetime it remains embedded in the progenitor molecular cloud until stellar winds, ionising flux from massive stars, and stellar feedback expel the remaining gas \citep[see][for a recent review]{goodwin08}. In expelling this gas many clusters become unbound and subsequently disperse into the surrounding medium \citep{lada03}, a process commonly described as 'infant mortality'. Even if clusters do survive the embedded phase intact other disruption mechanisms may come into effect, such as stellar evolutionary mass loss, two body relaxation and GMC encounters.  These mechanisms generally operate over longer time-scales, being a few 10s of Myr for stellar evolution (e.g. \citealt{bastian09}) and 100s of Myr for GMC encounters \citep{gieles06c}, except when the GMC number density is particularly high (e.g. \citealt{lamers05}).



After a cluster disrupts the stars disperse and become part of the galactic background. Comparing the fraction light emitted from clusters to the total (i.e. background plus clusters) galactic emission gives an indication of the fraction of stars within bound clusters. \citet{meurer95} estimated that $20-50\%$ of UV light in starburst galaxies comes from star clusters. \citet{zepf99} calculated the fraction of light from clusters in the B band to be $15-20\%$ and half that in the I band for the galaxy NGC~3256. However the most comprehensive set of results comes from \citet{larsen00}, listing the fraction of light from clusters in both the U-band (T$_L$(U)) and V-bands (T$_L$(V)) for 32 galaxies with varying star formation rates. \citet{larsen00} found that T$_L$(U) increases with the star formation rate of the host galaxy and a stronger correlation with the star formation rate surface density, indicating that the host environment may influence the mode of star and/or cluster formation and how likely a cluster is to survive. 

Measuring the fraction of light from clusters is a relative easy calculation assuming foreground stars can be eliminated from the sample. However, it is presently unknown, but calculable, how these values may be influenced by the presence of bursts of star formation in the past and differential extinction of younger clusters.   Hence the fraction of light observed in clusters will be a (possibly complicated) combination of the fraction of stars formed in clusters, the star-formation history of the galaxy, the cluster disruption time-scale/rate, and the difference in the amount of extinction towards clusters and the field. 

In this paper we attempt to improve the situation by deriving ages and masses for clusters in the galaxy NGC~3256, which in turn is used  to calculate a cluster formation rate (CFR) over the last ten million years. Comparing this inferred CFR to the star formation rate (SFR, measured by \ha fluxes or infrared luminosities) we compute the fraction of stars in clusters younger than ten million years which have survived the embedded phase intact, a value hereafter referred to as $\Gamma$ \citep{bastian08}. We pay particular attention to the possible sources of uncertainty which may affect these calculations. Using other data sets of cluster ages and masses for different galaxies we perform the same calculation for an additional 6 galaxies in an attempt to find how $\Gamma$ varies with environment. 

Throughout this paper we will define a "cluster" as a gravitationally bound, centrally concentrated group of stars that can be identified on high resolution optical imaging.  Hence, clusters refer to objects that have survived the transition from being embedded in their natal GMC to an exposed state.  We assume that this process happens approximately 3~Myr after the cluster forms. We adopt a Hubble constant of $H_{0} = 70$ km s$^{-1}$ Mpc$^{-1}$, which places NGC 3256 at a distance of 36.1 Mpc given it has a recession velocity relative to the local group of $+2804\pm6$ km s$^{-1}$. This corresponds to a distance modulus of 32.79.


This paper is structured as follows, we begin in \S~\ref{sec:obs} by introducing the dataset for NGC 3256 and the methods we used to determine the age and masses of clusters in the galaxy. In \S~\ref{sec:gamma_calc} we detail how we calculated $\Gamma$ for NGC 3256 and go on to carefully examine all the possible sources of error in making such a calculation. In \S~\ref{sec:results} we introduce other datasets taken from the literature and calculate $\Gamma$ for each.  In \S~\ref{sec:implications} we search for trends with $\Gamma$ and galactic properties and discuss the implication.  Finally, in \S~\ref{sec:conclusions} we present our conclusions and summarise our main results.


\section{Our Study of NGC 3256}
\label{sec:obs}


\subsection{The Data}

We used archival Hubble Space Telescope (HST) Advanced Camera for Surveys (ACS) images of NGC 3256 across four filters, {\it F330W} (PI: Holland Ford; ID: 9300), {\it F435W} (PI: Alex Fillippenko; ID: 10272), {\it F555W} (ID: 9300) and {\it F814W} (ID: 9300). Details of all the images used can be found in Table \ref{tab:ims}. The High Resolution Camera (HRC) images all cover the same area apart from the HRC-{\it F435W} image which is offset. To ensure we were examining the same area in all four bands we used the ACS Wide Field Camera (WFC) {\it F435W} (PI: Aaron S. Evans; ID: 10592) image over the HRC {\it F435W} image. The WFC image has the advantage of a longer exposure time, though the pixel scale is slightly larger, $0.05\arcsec$ compared to $0.027\arcsec$. Throughout this paper we will refer to the {\it F330W}, {\it F435W}, {\it F555W}, and {\it F814W} in the Cousins-Johnson {\it U}, {\it B}, {\it V} and {\it I} notation, although we stress that no transformations have been applied.

\begin{table}
\caption{Details of images}
\label{tab:ims}
\begin{tabular}{@{}l c c c c}
\hline
\hline
Filter & Camera & Exposure & Aperture & Name$^a$ \\
	& 	& Time (s) 	& Correction	& \\
\hline
F330W & HRC & 2000 & 0.418 & U \\
F435W & HRC & 840 & - & - \\
F555W & HRC & 760 & 0.397 & V \\
F814W & HRC & 660 & 0.405 & I \\
\hline
F435W & WFC & 1320 & 0.328 & B \\
\hline
\end{tabular}

$^a$ We refer to the filters by their closest Cousins-Johnson filter name, however no transformations have been applied.
\end{table}

\subsection{Cluster Selection \& Photometry}

We selected star clusters using the \sex program \citep{bertin96}, run on the B-band WFC image in order to maximise the number of clusters detected. We changed the \sex detection parameters to maximise the number of objects found, attempting to avoid blending of objects in the most crowded regions. We set a minimum detection limit as $4\sigma$ above the local background as determined by \sex, and a minimum object area of 8 pixels. Co-ordinates of the clusters were translated from the B band image into co-ordinates for the other bands using the \emph{GEOMAP} and \emph{GEOXYTRAN} routines within \emph{IRAF}. 

Photometry across all images was carried out using the \emph{APER} routine from the \emph{DAOPHOT} package in \emph{IRAF}. We did not use the photometry from \sex to ensure a consistent and unbiased method in all wavebands. We used an aperture of radius $7.0$ pixels for the HRC images and $3.5$ pixels for the WFC image. An annulus  with radius 8 pixels and a thickness of one pixel was used to measure the local background in the HRC images, radii and thickness of the annulus were adjusted accordingly for the WFC image. Aperture corrections were calculated based on several bright, isolated and spatially resolved clusters, by comparing their flux in an aperture of 30 (HRC) pixels relative to our adopted 7 (HRC) pixel aperture. The resulting aperture corrections are shown in Table \ref{tab:ims}.

We corrected the photometry for  foreground galactic extinction, using the extinction law of \citet{savage79}, and a dust extinction value of  A$_{V} = 0.122$, taken from \citet{schlegel98}.

\begin{figure}
   \includegraphics[width= 84mm]{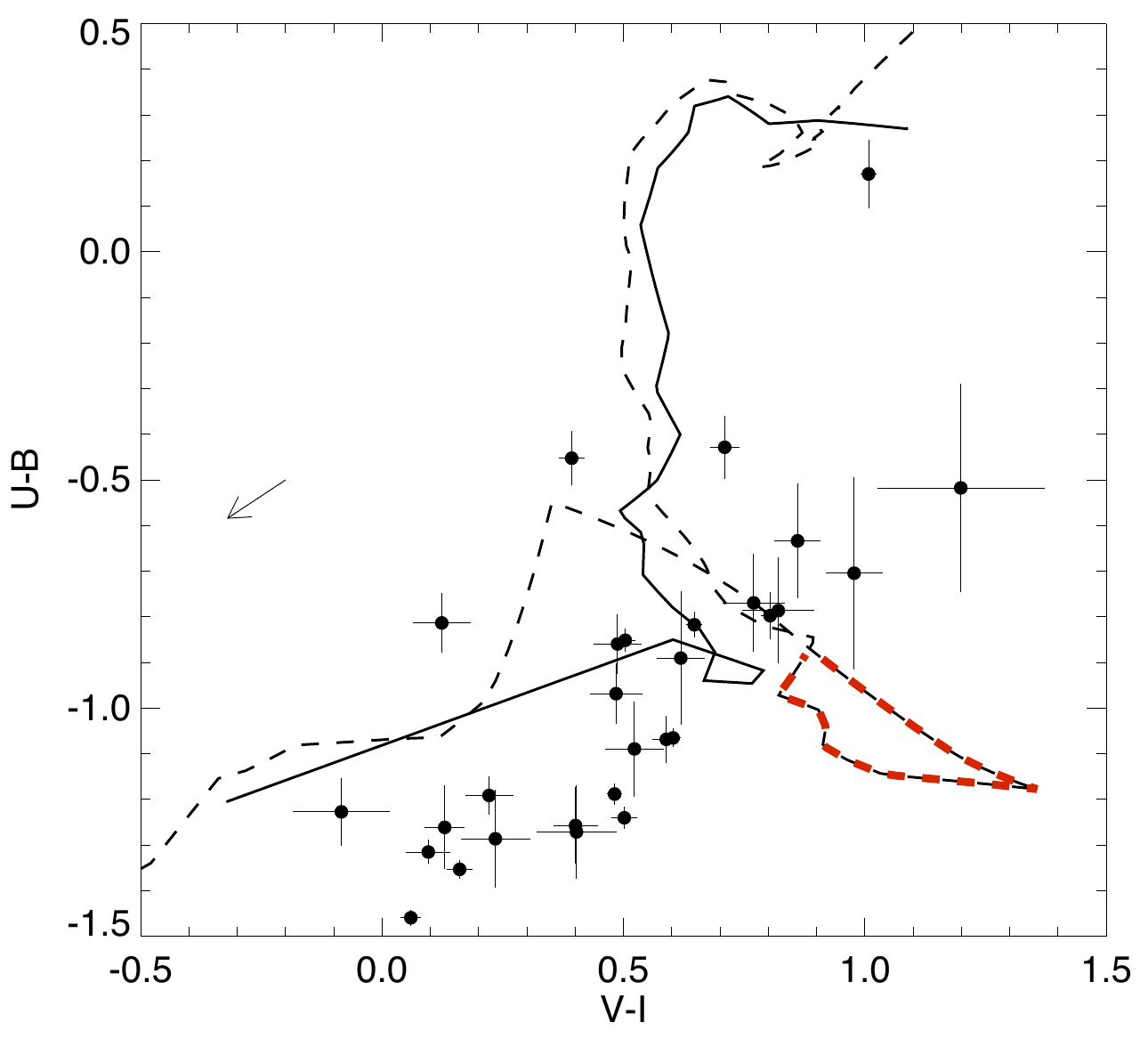}
\caption{Colour-Colour diagram for clusters below the magnitude limit $m_B < 21.5$ (simply to avoid overcrowding the figure). Clusters have not been reddening corrected. We have only included clusters that met our criteria for a \emph{good} fit to the SSP models, as explained in section \ref{sec:good}. The lines show solar metallicity SSP models, the solid line is based on the Padova isochrones whilst the dashed line is based on the Geneva isochrones. The red dashed segment shows the 'red-loop', few clusters are ever associated with this feature.}
\label{fig:ubvi}
\end{figure}
\begin{figure}
   \includegraphics[width= 84mm]{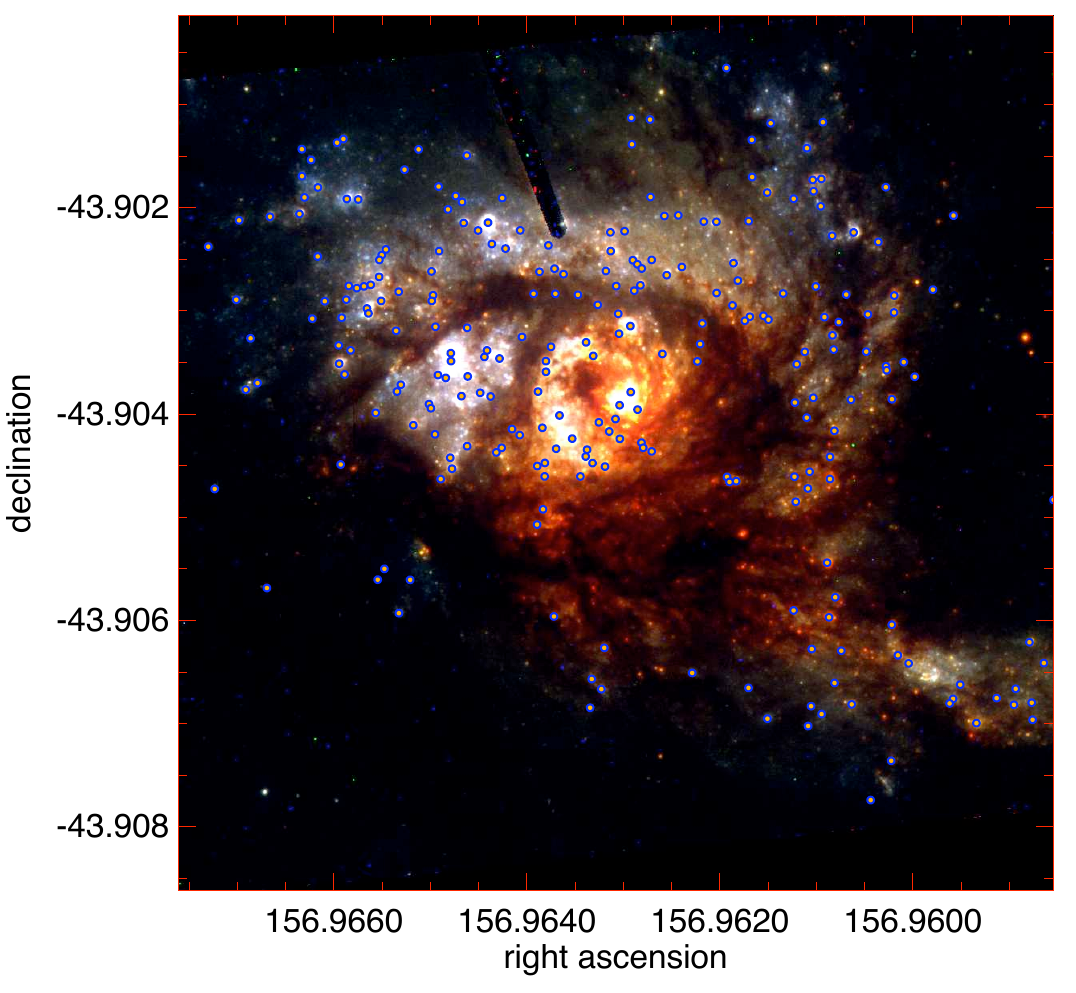}

\caption{Colour composite image of of NGC 3256. We have marked on the positions of the 276 clusters classified as \emph{good} fits in section \ref{sec:good}.}
   \label{fig:galim}
\end{figure}

In total we measured fluxes for 904 clusters in NGC 3256 across the U, B, V and I bands. Figure \ref{fig:ubvi} shows a colour-colour diagram of clusters within a magnitude limit ($m_B < 21.0$). Figure \ref{fig:galim} shows a colour-composite image of NGC 3256. To avoid overcrowding we have only marked the positions of the 276 cluster that have \emph{good} fits to SSP models, more details on this selection can be found in section \ref{sec:good}.

\subsection{Determination of Cluster Parameters} 
\label{sec:good}

In order to determine the age, mass and local extinction for each cluster the photometric values were compared to those of cluster evolution models, in a similar fashion to \citet{bik03} and \citet{bastian05}.  We used the most recent GALEV simple stellar population (SSP) models  \citep{anders03,anders04}. Although SSP models exist based on both the Padova and Geneva isochrones, we have only used the Padova tracks to determine cluster parameters. The Geneva tracks display a feature known as the red loop and tend to produce a poor fit to data compared to the Padova tracks \citep{whitmore02}. The GALEV models have the advantage of being produced with colours matching the HST filters, this avoids errors associated with converting between HST filters and the standard Cousins-Johnson system. In addition, the GALEV models include gaseous emission lines and continuum emission. The youngest data points in the GALEV models are 4Myr and 8Myr, this can lead to poor fitting of the youngest clusters \citep{bastian05}. To avoid this we linearly interpolate between the youngest model ages for all filters. This produces SSP tracks which are regularly sampled at intervals of 0.02 in log time. The determined age of a young cluster is not bound to 4 or 8Myr in this case and can take a more precise value with a better fit over the four wavebands.

We have used a three-dimensional maximum likehood fitting method developed by \citet{bik03}. The method is described in detail there, and so we only summarise the process in this paper. We also note that this method has been tested against colour-colour methods and has shown itself to be superior \citep{degrijs03a, parmentier03}.

In simple terms the method works as follows: the GALEV SSP models give a grid of colours as a function of age. We then apply extinction to each model in steps of 0.02 in E(B-V) to extend the grid. Each cluster then has its observed spectral energy distribution compared to the grid using a minimum $\chi^2$ test. The model with the age and extinction which produces the lowest $\chi^2$ is chosen, and a range of acceptable values is calculated. As the GALEV model magnitudes are scaled to a mass of $10^6$\msol \ the difference between the observed magnitudes and model magnitudes can be converted to a mass, given the distance modulus and extinction. The accuracy of this fitting method is not studied directly here as it has been reported previously \citep{bastian05}, for the majority of the sample errors on the age and mass are less than 0.3 dex.

After obtaining the best fit for each cluster we sought to determine if the fit was good or not. In fitting cluster parameters a $\chi^2$ value is calculated and this is one way of identifying a \emph{good} fit. However for bright objects with low photometric errors the $\chi^2$ value may be high and thus result in a \emph{bad} fit. To avoid such instances we use the standard deviation, defined in Equation \ref{eqn:stddev}.

\begin{equation}
\label{eqn:stddev}
\sigma^2 = \sum\limits_i \frac{(m_i^{mod} - m_i^{obs})^2}{n_{filters}}
\end{equation}

The $\sigma^2$ value is summed over all four wavebands. The spectral energy distributions were compared to the values predicted by the best fit models and compared by eye. This was done to establish a minimum value for $\sigma^2$, $\sigma_{min}^2$. We also computed a histograms of $\sigma^2$ for all sources and approximating the distribution with a gaussian and took the $\frac{1}{e}$ point.  This value was very close to an appropriate value for $\sigma_{min}^2$ determined by visual inspection at 0.03.

In the resulting analysis of the properties of cluster system of NGC 3256 we only consider those clusters which pass the following criteria:

\begin{enumerate}
\item  detected in all four wavebands, at least 5$\sigma$ above the background;
\item uncertainties in the magnitude of all wavebands $\le$ 0.2 mag;
\item is well fitted by an SSP model ($\sigma \le 0.03$).
\end{enumerate}

The number of clusters which passed these rigourous selection criteria depends on the metallicity of the SSP model used. With a solar metallicity Padova model we accept 276 objects, with a twice solar metallicity we accept 285 objects. Details of the 276 clusters accepted using the solar metallicity Padova tracks can found with the electronic version of this article. It should be noted that the higher number of accepted objects produced by the twice solar metallicity models is not an indication of the global metallicity of the galaxy. Metallicity is very hard to fit accurately with only photometry, as shown by \citet{bastian05}. \citet{trancho07a} found the metallicity to range between $1.1 - 1.7\ $\zsol \ for 23 clusters in NGC 3256.

\citet{trancho07a} (hereafter T07a) presented optical spectroscopy of 23 clusters in NGC3256. We have attempted to compare our results with those of T07a, however there are several problems that we encountered. Firstly, the astrometry presented by T07a is not accurate enough to identify any one cluster from our study, we were forced to visually inspect images from the two studies and then attempted to find the most likely matching clusters. Secondly, several of the clusters presented by T07a are in fact complexes, comprised of several clusters. Out of the ten clusters in the T07a study that lie within the same field of view as our study we found that eight had corresponding clusters in our study which matched our criteria for a \emph{good} cluster fit. The remaining two clusters from the T07a were both complexes located in extremely crowded regions of the galaxy and could correspond to any one of several clusters in our study. Of the eight matched clusters T07a states that six of these were emission line objects and have reported ages of less than $10^{6.8}$ years. We find that ages for seven of these clusters is also under $10^{6.8}$ years, the remaining object has an age between $10^{6.7} - 10^{6.9}$ which appears to be consistent with the findings of T07a. We have also be able to match two absorption line clusters from the T07a study, both of these clusters have reported ages that are consistent within their respective errors. 


\subsection{Cluster Properties of NGC 3256}\label{sec:clprop}

\begin{figure}
   \includegraphics[width= 84mm]{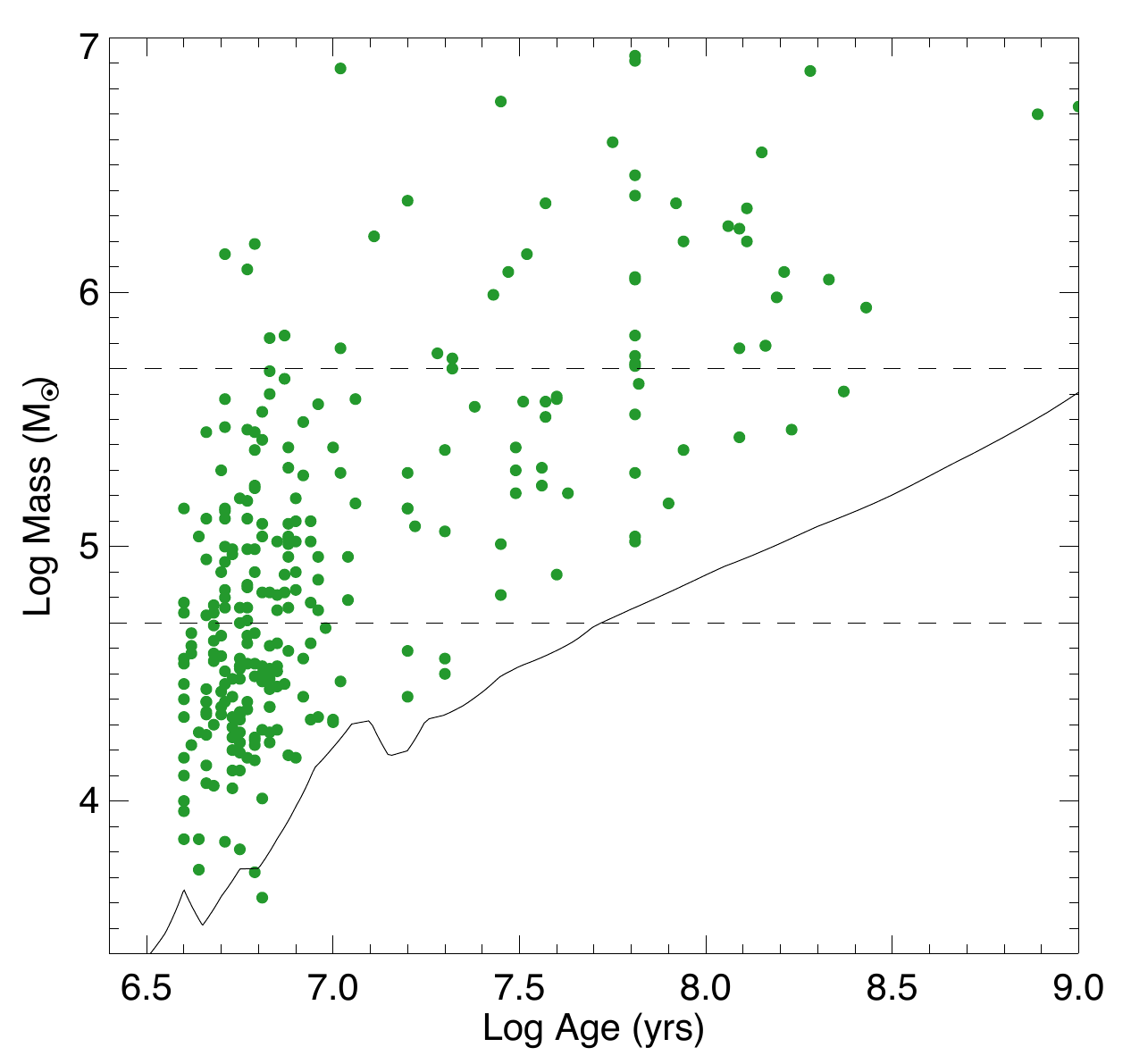}

\caption{The age-mass distribution for all clusters with a \emph{good} fit, using solar metallicity Padova SSP models. The dashed lines represent mass cuts at 4.7 and 5.7 log(\msol), which are used in figure \ref{fig:hist} and throughout the rest of the study. The solid black line represents the magnitude limit m$_{V} = 23.5$ based on the solar metallicity Geneva SSP models at the distance of NGC~3256}
\label{fig:agemass}   
\end{figure}

We can determine the properties and history of NGC 3256 by examining the clusters that reside within the galaxy. With the age, mass and extinction of the clusters known we can derive several important properties. Figure \ref{fig:agemass} shows the age/mass distribution for NGC 3256 derived from the solar metallicity Padova SSP models. Immediately obvious from Figure \ref{fig:agemass} is the lack of low mass, old clusters. This is due to the fact that clusters dim as they age and so eventually become fainter than our detection limit. We can also note the large number of clusters seen with ages below 10Myr and over a large range of masses. Although the cluster fitting method can create some observed structure in the age/mass diagram \citep{gieles05} it is unlikely to do so over all masses at low ages. We can conclude that there is a genuine over-density of clusters with ages below 10Myr. This can be interpreted as a lack of older clusters which could be due in part to the disruption of clusters, a point we shall address when we look as the cluster formation history in \S~\ref{sec:formation-history}.

Various mass limits are shown on Figure \ref{fig:agemass} as horizontal dashed lines. Due to the effects of old clusters becoming fainter, higher mass limits result in a population which is complete up to a larger range of cluster ages. For example a mass limit of $log(M/\msun)>4.5$ would only be complete for clusters younger than 10Myr. A mass limit of $log(M/\msun)>5.5$ would however be complete for much older clusters, those less than $\sim200$~Myr.

\begin{figure}
   \includegraphics[width= 84mm]{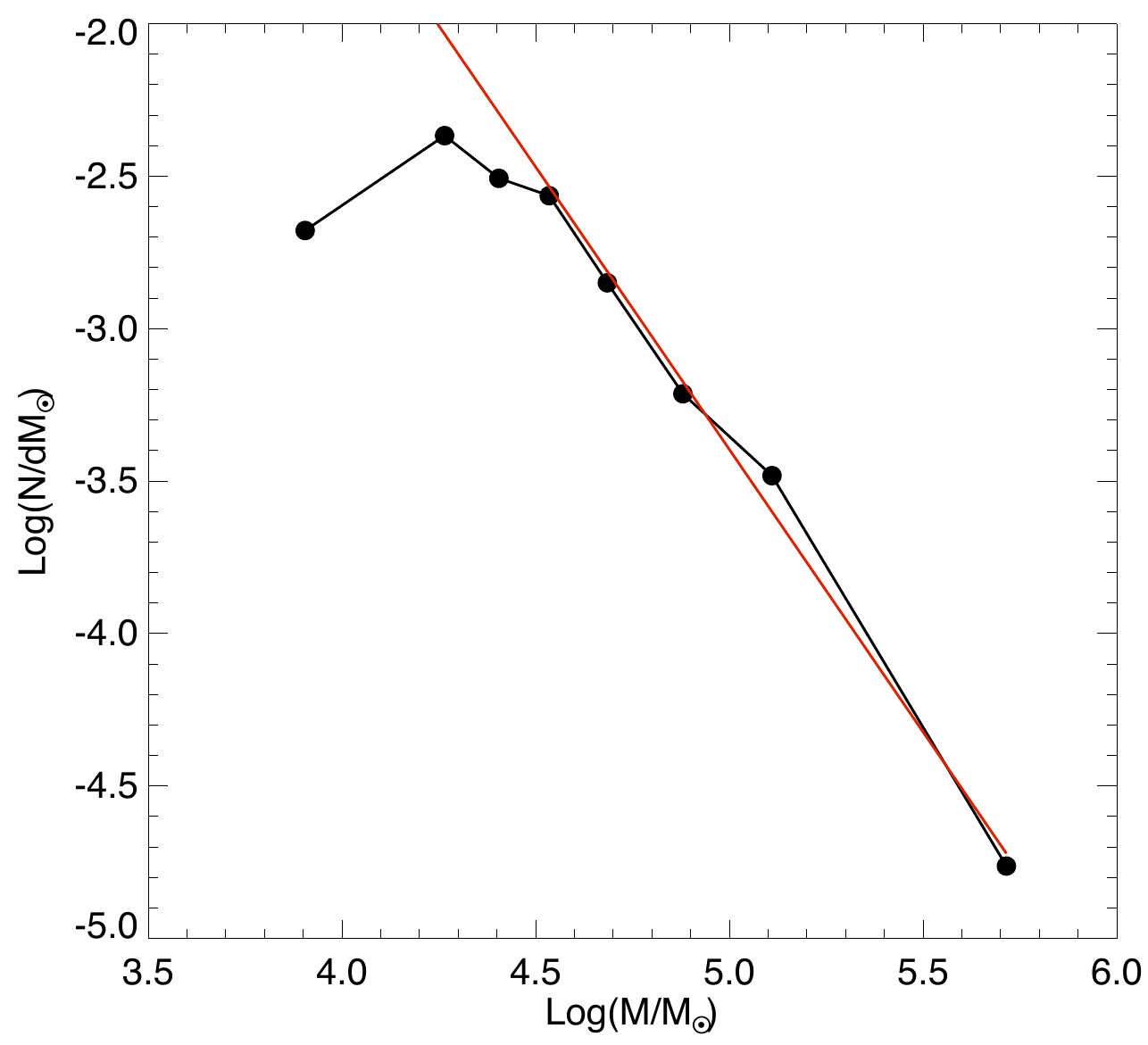}

\caption{Mass function for clusters younger than 10 million years old in NGC 3256, produced using bins with a constant number of clusters per mass bin. The red line is a line of best fit for points beyond the low mass turn over.}
   \label{fig:massfn}
\end{figure}

We can compute the slope of the cluster mass function for clusters younger than 10 million years which we have shown in Figure \ref{fig:massfn}. This assumes the number of clusters of a given mass follows a power law distribution, like that in Equation \ref{eqn:mass}. $n(M)$ is the number of clusters with mass $M$, $\chi$ is a constant of normalisation and $\alpha$ is the slope of the power law.

\begin{equation}
\label{eqn:mass}
N(m)dm = \chi M^{-\alpha}dm
\end{equation}

When calculating the slope of a mass distribution for any type of object the way in which the data are binned can affect the result \citep{maiz05}. To avoid such biases we have used bins with an equal number of objects in each. The minimum and maximum limits to each bin then depend on the masses of objects within. We have chosen to calculate the mass function for young clusters with ages less than 10Myr, in order to avoid any mass dependent cluster disruption effects, resulting in a complete sample.

This calculation produces the mass function shown in Figure \ref{fig:massfn} where the data points are shown in black and the line of best fit is in red. The slope is fit to data points with masses above the limit $M/\msol > log(4.3)$, below which we see a turn over in the mass function. This common turn over in cluster mass functions is due to incompleteness in the sample for low mass clusters. The slope of the mass function for NGC 3256 was measured to be $\alpha = 1.85\pm0.12$, which is similar to other observed cluster populations that are generally in the range of $\alpha = 1.8 - 2.2$  \citep{degrijs03c, mccrady07}.

\subsection{Cluster Formation History of NGC 3256}
\label{sec:formation-history}

Ages determined by fitting photometric observations to SSP models can lead to large uncertainties, especially for faint objects with large photometric errors. In constructing a cluster formation history for NGC 3256 it is important to take the uncertainty in the cluster ages into account.

We have constructed a $\frac{dN}{d\tau}$ distribution using the same method described by \citet{gieles07}. We do not reiterate the details laid out in \citet{gieles07}, only the central ideas. Each cluster has a contribution to the overall age distribution as defined by an asymmetrical Gaussian spread in $log\ t$. This is calculated based on the minimum and maximum allowed ages of each cluster, calculated by the SSP fitting routine.

\begin{figure*}
   \includegraphics[width= 84mm]{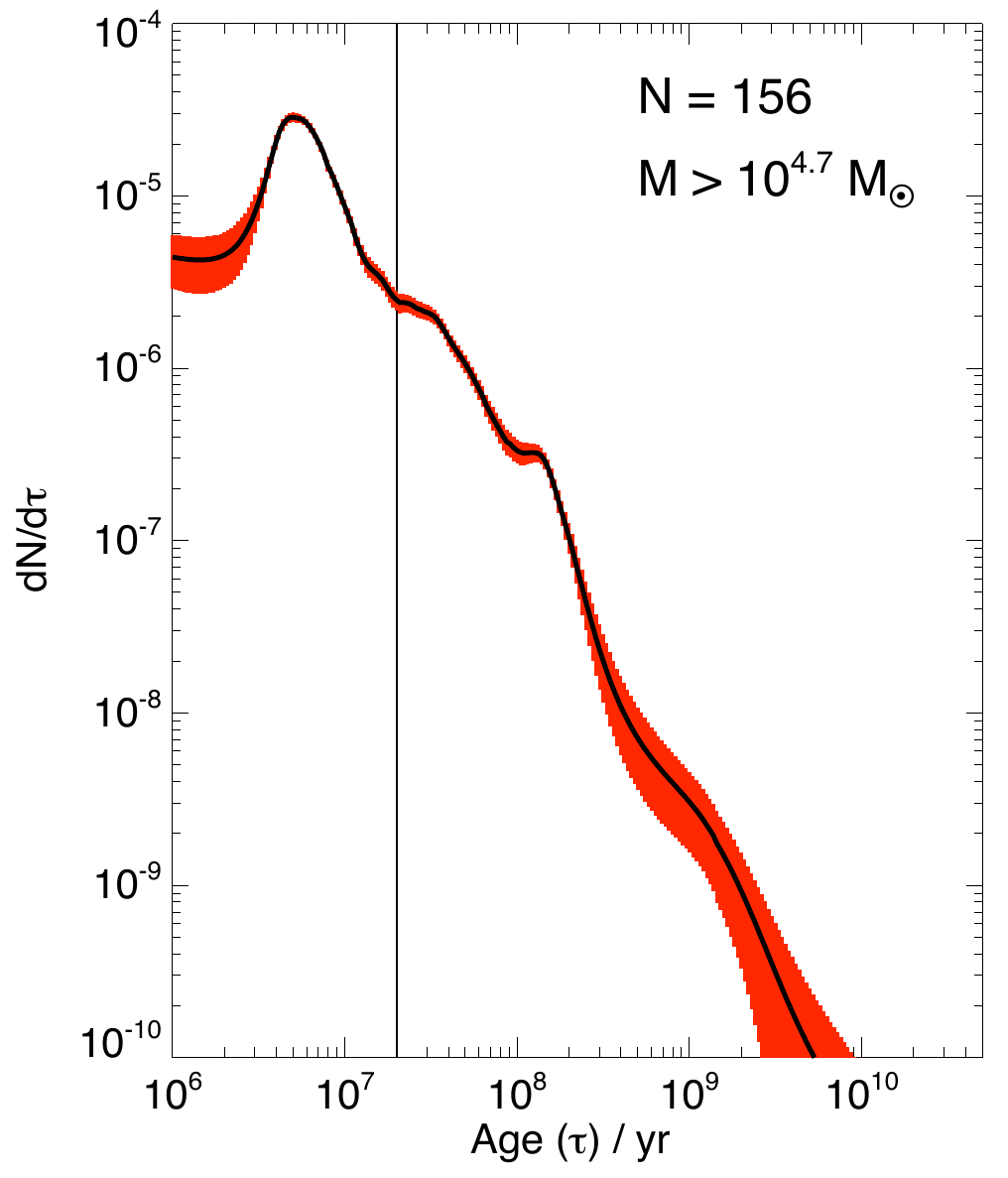}
      \includegraphics[width= 84mm]{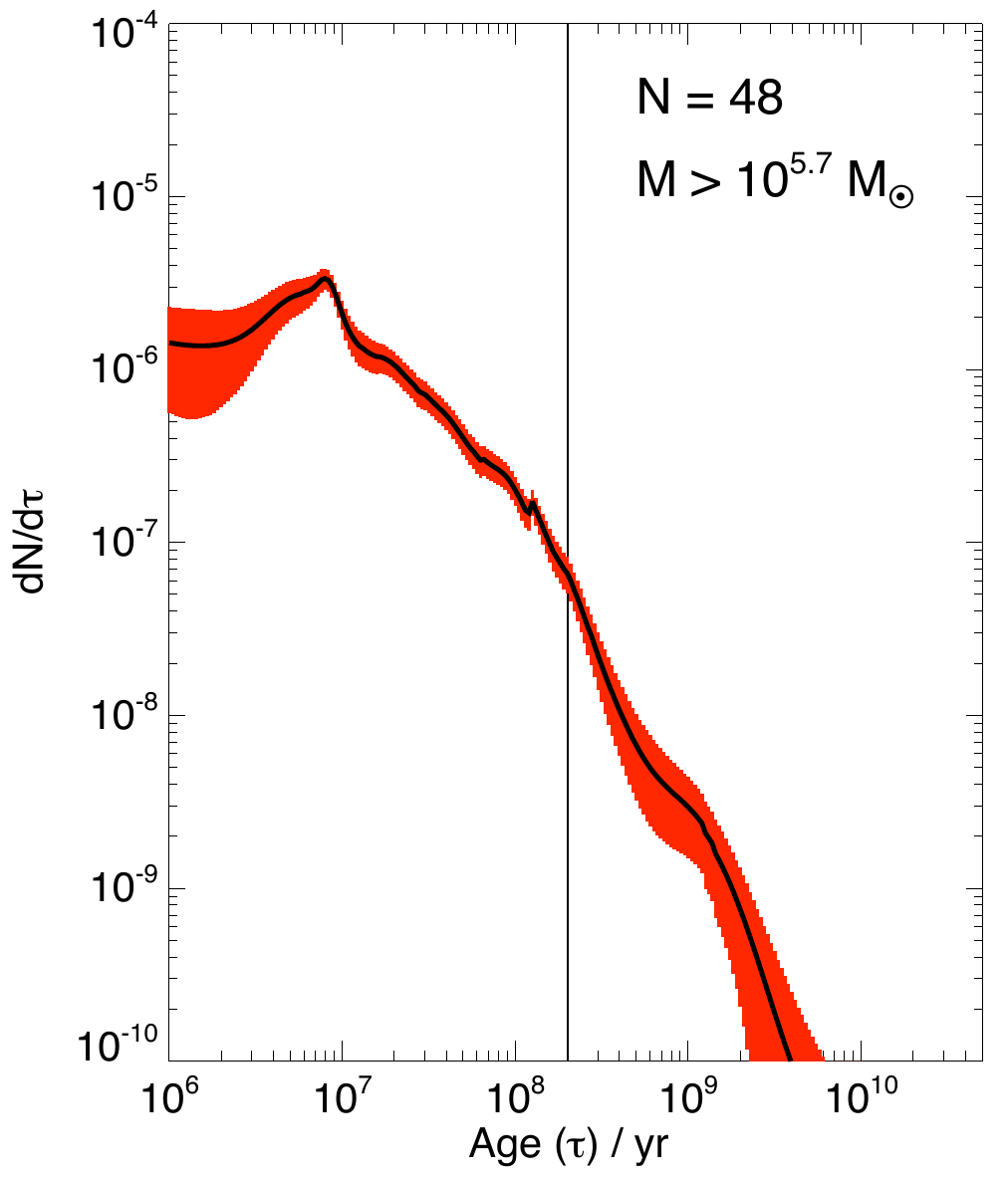}

\caption{Apparent rate of cluster formation as a function of time for NGC 3256 and 2 different lower cluster mass limits. Left panel uses a limit of $10^{4.7}$ \msol\ which is complete for clusters younger than 20 million years, represented by the solid vertical line. The right panel shows the same plot for clusters more massive than $10^{5.7}$ \msol, which is complete up to 200 million years. The red shaded area denotes the $1\sigma$ Poisson errors as described in the text.}
   \label{fig:hist}
\end{figure*}

In Figure \ref{fig:hist} we show the smoothed age distribution results for two different lower mass limits, $10^{4.7}$ \msol\ (left panel) and $10^{5.7}$ \msol\ (right). The red shaded area shows the $1\sigma$ Poisson errors, estimated by counting the number of clusters in bins of width 0.25, corresponding to the mean uncertainty in the log age values of the clusters. 

Using two different lower mass limits in Figure \ref{fig:hist} our sample of clusters is complete up to two ages, the $10^{4.7}$ \msol\ limit shown in the left panel is complete for clusters younger than 20 million years. The higher mass limit of $10^{5.7}$ \msol\ shown on the right is complete for clusters younger than 200 million years, both of these ages are shown as a vertical line on each plot. In both panels the cluster formation rate declines for clusters older than the completeness limit, due in part to older clusters fading and becoming undetectable.

Focussing on the high mass cut sample, we see that the cluster formation rate (CFR) has increased by a factor of $\sim10$ during the past $\sim200$~Myr.  This increase is expected due to the ongoing galactic merger which is presumably the cause of the ongoing starburst \citep{bastian09}.  This interpretation is also supported by the present SFR of NGC~3256 which is $\sim46$~\msunyr, which is approximately ten times higher than the sum of the two progenitor spiral galaxies.  As in the Antennae \citep{bastian09}, we do not see evidence for a high degree of long duration ($>10$~Myr) mass independent cluster disruption, as proposed by \citet{fall05}.  However, our high mass cuts means that we are insensitive to any disruption in the lower mass end of the cluster mass function, which may be strongly effected by cluster disruption during galaxy mergers (Kruijssen et al.~2010 in prep.).

\section{Calculating the Ratio of Stars Forming Within Clusters, $\Gamma$}
\label{sec:gamma_calc}

\subsection{Initial Estimate}

Assuming we were able to detect all the clusters within a galaxy we would be able to establish the total mass of clusters which had recently formed. Comparing this mass with a measure of the star formation rate (SFR) over the same time scale gives the ratio of stars forming within clusters, hereafter referred to as $\Gamma$ \citep{bastian08}.

In principle this calculation is simple but is made more complex by the incompleteness of our sample of clusters. We are not able to detect low mass clusters, even those with young ages. Any low mass cluster is also likely to have large photometric errors and thus be rejected as a \emph{poor} fit. We know, however, our sample is likely to be complete for bright, massive, young clusters. We can then extrapolate the mass found in these clusters to calculate the total mass of all clusters.

In order to calculate the ratio of mass found above a certain mass limit we made sample cluster populations based on a power law distribution (stochastically sampled), like that in Equation \ref{eqn:mass}. We made model populations with 500,000 clusters with a lower mass limit of 100\msol, and an upper limit of $10^{11}$\msol. This upper mass limit may appear very high but with a power law distribution with index of $-2.0$ (i.e. $\alpha=2.0$) we only expect 50 clusters above $10^6$\msol\ based on this upper mass limit.   To avoid erroneous results we ignore clusters with masses larger than $10^7$\msol. The upper mass limit allows the model populations to be fully populated for high masses. We also calculate an error associated with the fraction of mass found above a certain mass limit, this is the standard deviation of results from 200 model cluster populations. The error does not decrease with an increased number of model populations as it is limited by stochastic variations in the masses of high mass clusters, which are rare in each model population. We have used a power law index of $-2.0$ in our claculations although it should be noted that an index of $-1.85$ was measured for NGC 3256.  We assume an index of $-2.0$ as this is the value found for large samples of clusters, we do accept that this value may vary between galaxies and the implications of this are discussed further in \S~\ref{sec:uncertainty}.

At this point we note that sampling an ICMF to generate a cluster population simply to calculate the mass above a mass limit may appear to be a long winded way of making this calculation, as it is straightforward to analytically calculate these numbers for a given ICMF. However analytical calculations do not give an estimate of the error associated with stochastic sampling of the ICMF, an effect that increases as you estimate the fraction of mass above an increasing mass limit. As estimating realistic errors for this and similar studies is one of the goals of this study, we choose to use the monte carlo technique in the subsequent analysis.

With the ages and mass of clusters for NGC 3256 calculated we found the total mass of clusters less than 10 million years old, and with a mass greater than $10^{4.7}$\msol. The lower mass limit of $10^{4.7}$\msol\ was chosen as a conservative estimate to ensure our sample was complete for this range of masses and ages. This gave a mass of $1.66\times 10^7$\msol\ contained within clusters. The fraction of mass expected above this mass limit was taken from our model cluster populations to be $0.61\pm0.09$. We expect the total mass contained within all clusters in this age range to be $2.73\pm0.42\times10^7$\msol. This was divided by the age range of 7 million years to find the stellar mass forming in clusters per year, $3.90\pm0.60$\msol\ yr$^{-1}$. We used an age of 7 rather than 10 million years as extremely young clusters with ages less than 3 million years will still be embedded, an thus undetectable.

We compare the mass of stars forming within clusters to the SFR which was calculated based on the IR luminosity from \citet{sargent89} and converted to a SFR of $46.17$ \msol\ yr$^{-1}$ using the conversion of \citet{kennicutt98}. The conversions given by \citet{kennicutt98} are based on a Salpeter IMF. In fitting SSP models to our photometric data we used SSP models which adopt a Kroupa IMF. This has the effect of underestimating the mass of clusters compared to a similar result using SSP models based on a Salpeter IMF. If we are to compare the mass of stars in clusters in clusters to the SFR both results must be made using the same IMF. We correct the masses of clusters for this by multiplying by 1.38.

For NGC~3256, we find that the fraction of stars which remain within optically selected clusters after the embedded phase, $\Gamma$, is $0.12\pm0.02$. This is, however, a lower limit as various selection effects and fitting artefacts must be taken into account, which will be explored below. The implications of this result are discussed in \S~\ref{sec:implications}.

\subsection{Sources of Uncertainty} 
\label{sec:uncertainty}

In calculating $\Gamma$ there are several sources of error we should consider. Here we focus on the errors associated with our study of NGC 3256 and also address issues which may arise with other data sets from different galaxies.

\subsubsection{Error Associated with the slope of the mass function}\label{sec:sloperr}

\begin{figure}
   \includegraphics[width= 84mm]{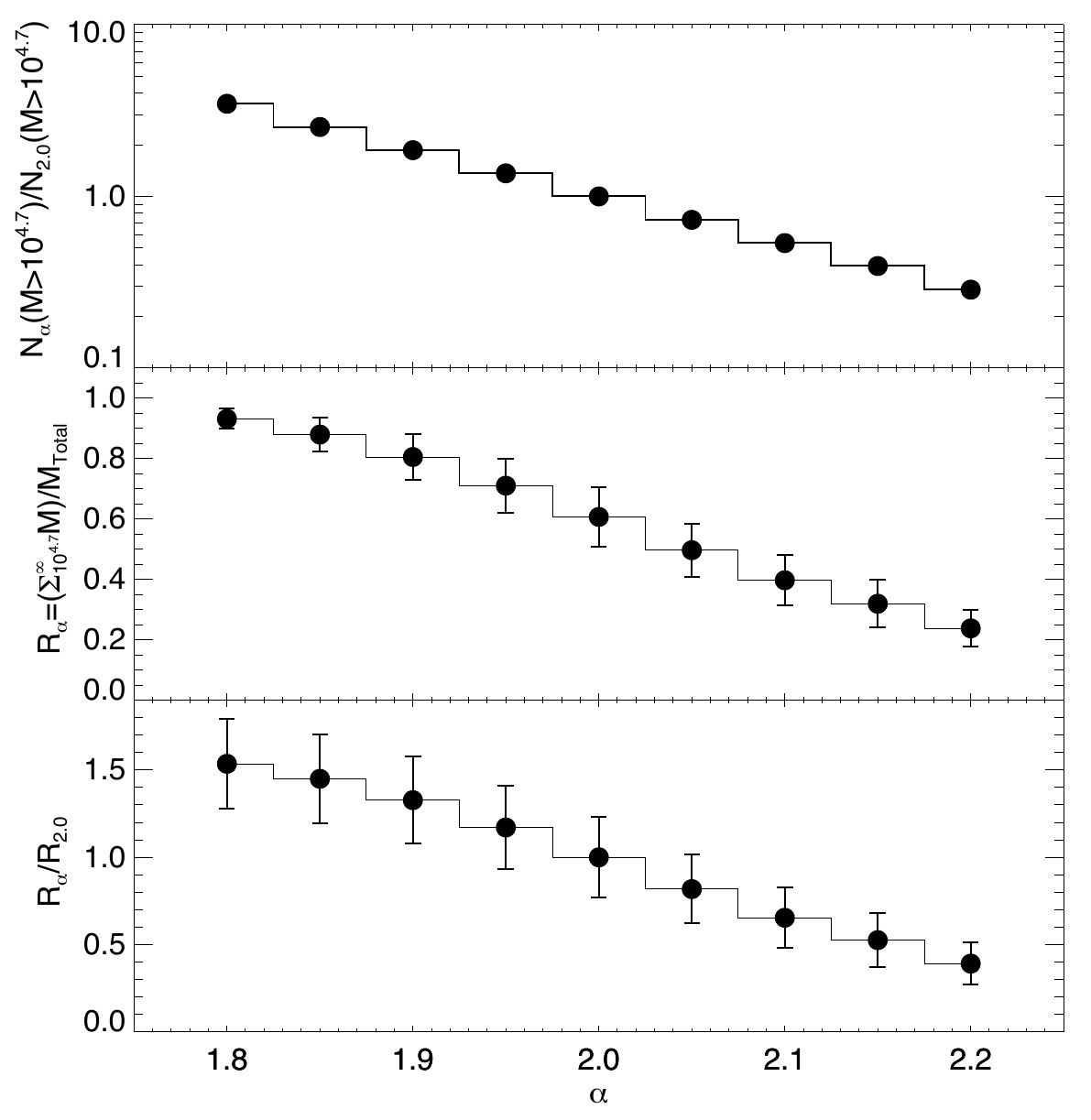}

\caption{Plot showing the effect of varying the mass function index on the fraction of mass above a mass limit ($M>10^{4.7}$). The top panel shows the number of clusters found above the mass limit as a function of the ICMF index ($-\alpha$), normalised to the value for a slope of $\alpha = 2.0$. The middle panels displays $R_\alpha$, the ratio of mass found in clusters above the mass limit to the total mass of all clusters. The bottom panel shows $R_\alpha$ normalised to $R_{2.0}$. }
   \label{fig:sloperr}
\end{figure}

As we previously mentioned, the slope of the cluster mass function was taken to be $-2.0$. However this may vary between galaxies and has an effect on the fraction of mass we infer to be above the mass limit. With a steeper slope fewer high mass clusters are present so the fraction of mass above a mass limit is reduced. Conversely a shallower slope creates more high mass clusters and so the fraction of mass above a mass limit is increased.  We have measured this effect by making sample cluster populations with a mass function slope between $-1.8$ and $-2.2$. The results of these simulations are shown in Figure \ref{fig:sloperr}, in which we assume a mass cut of $M>10^{4.7}$, the same used in our study of NGC 3256.  The top panel in Figure \ref{fig:sloperr} shows how the number of objects found above the mass cut varies with differing mass function slopes.

The middle panel of Figure \ref{fig:sloperr} shows the ratio of mass within clusters with masses greater than the mass limit compared to the total mass of all the clusters, $R_{\alpha}$.  Over the range of $1.8 < \alpha < 2.2$ this ratio changes dramatically from 0.93 to 0.25 respectively. Measured values of $\alpha$ vary but the best estimates are close to $2.0$. The bottom panel in Figure \ref{fig:sloperr} shows how the value of $R_{\alpha}$ compares to the value for a mass function with an index of $-2.0$. If the cluster IMF index varies between $1.8 < \alpha < 2.2$ then we may be under or over estimating $\Gamma$ by roughly $\pm50\%$ by assuming a single value of $\alpha=2.0$. 

\subsubsection{Power Law or Schechter Function?}

The exact form of the initial cluster mass function (ICMF) is still under debate. In the case of NGC 3256 we have assumed a simple power law of the form shown in Equation \ref{eqn:mass}. For other galaxies an index of $\alpha \approx 2$ has been found (e.g. \citealt{zhang99,degrijs03b,bik03,mccrady07}) and is expected from theoretical considerations \citep{elmegreen97}. Recent studies have, however, shown that there may be truncation in the mass function at the high mass end \citep{gieles06b,bastian08, larsen09}.  This can be represented by a Schechter function \citep{schechter76} which behaves as a pure power law in the low mass regime and has an exponential fall off above a given value, \mstar. 

\begin{figure}
   \includegraphics[width= 84mm]{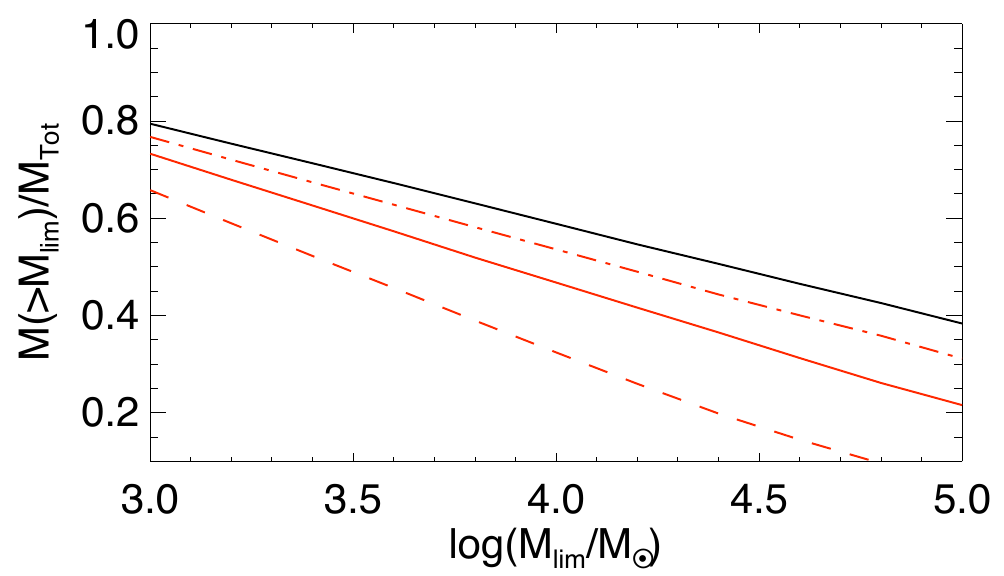}

\caption{The fraction of mass in clusters above a mass limit (M$_{lim}$) for an ICMF based on a Schechter function with $\beta = 2.0$ and  variable M$_*$ value. The black line shows the same fraction for a power law ICMF with identical slope, as well as three Schechter functions with differing M$_*$ values (dashed line, $10^{5.2} \msol$;  solid line, $10^{6.0} \msol$;  dot-dashed line, $10^{6.8} \msol$) }
   \label{fig:schec}
\end{figure}

In Fig.~\ref{fig:schec} we show how the resulting fraction of mass in clusters varies for different mass cuts and different truncation values, \mstar.  Overall, the difference between the results with an underlying power-law or a Schechter function are small.  In the specific case of NGC~3256, if a truncation exists it is likely to be well above $10^{6}\msol$, hence any effect on the derived $\Gamma$ values will be minimal.



\subsubsection{Error Associated with Fitting Cluster Parameters using SSP models} \label{sec:errs1}

\citet{bastian05} and \citet{gieles05} compared the results of SSP fitting for clusters in M51 with the results for a model population, noting that the number of clusters at certain ages and masses can be enhanced or reduced by the fitting method. This can affect the calculated total mass of clusters when extrapolating from the mass found in high mass clusters. To investigate this we made a model population of clusters based on the solar metallicity {\it GALEV} SSP model tracks (Padova isochrones). Ages of clusters were randomly distributed in the range $0 < t < 10^{10}$ yrs. The masses of clusters were assigned stochastically assuming a power law distribution of slope $-2.0$. Based on the age, masss and distance to NGC~3256 we then assigned magnitudes from the SSP models.

Photometric errors for each band were estimated by examining the errors of the clusters measured in our sample from NGC 3256. We were able to approximate the error with the relation $\Delta m_{\nu} = 10^{d_1+d_2\times m_{\nu}}$. The values of $d_1$ and $d_2$ are displayed in Table \ref{tab:errs}. A random correction to the magnitude in each waveband was added in the range $-\Delta M_{\nu}$ to $+\Delta M_{\nu}$. We produced a catalogue of 2607 clusters which fulfilled our magnitude limit.

\begin{table}
\caption{Parameters for the uncertainty of the magnitudes: $\Delta m_{\nu} = 10^{d_1+d_2\times m_{\nu}}$.}
\label{tab:errs}
\begin{tabular}{@{}l c c}
\hline
\hline
Filter & $d_1$ & $d_2$ \\
\hline
U & -5.726 & 0.228 \\
B & -3.276 & 0.116 \\
V & -4.881 & 0.178 \\
I  & -5.134 & 0.196 \\
\hline
\end{tabular}
\end{table}

We ran the same fitting procedure on this artificial catalogue as we had done on the data for NGC 3256. Fitting cluster photometry to the solar metallicty Padova SSP models. When examining the results of this fitting procedure we were as stringent with the artificial clusters as we were with the real data. We only accepted those results which met our criteria to be a \emph{good} fit, as defined in \S~\ref{sec:good}. We then compared the known mass, age and extinction of the model population to the results from the model fitting. We also excluded those model clusters with ages less than 3 million years, assuming that in reality these young clusters would be embedded and unobservable. 

\begin{figure}
   \includegraphics[width= 84mm]{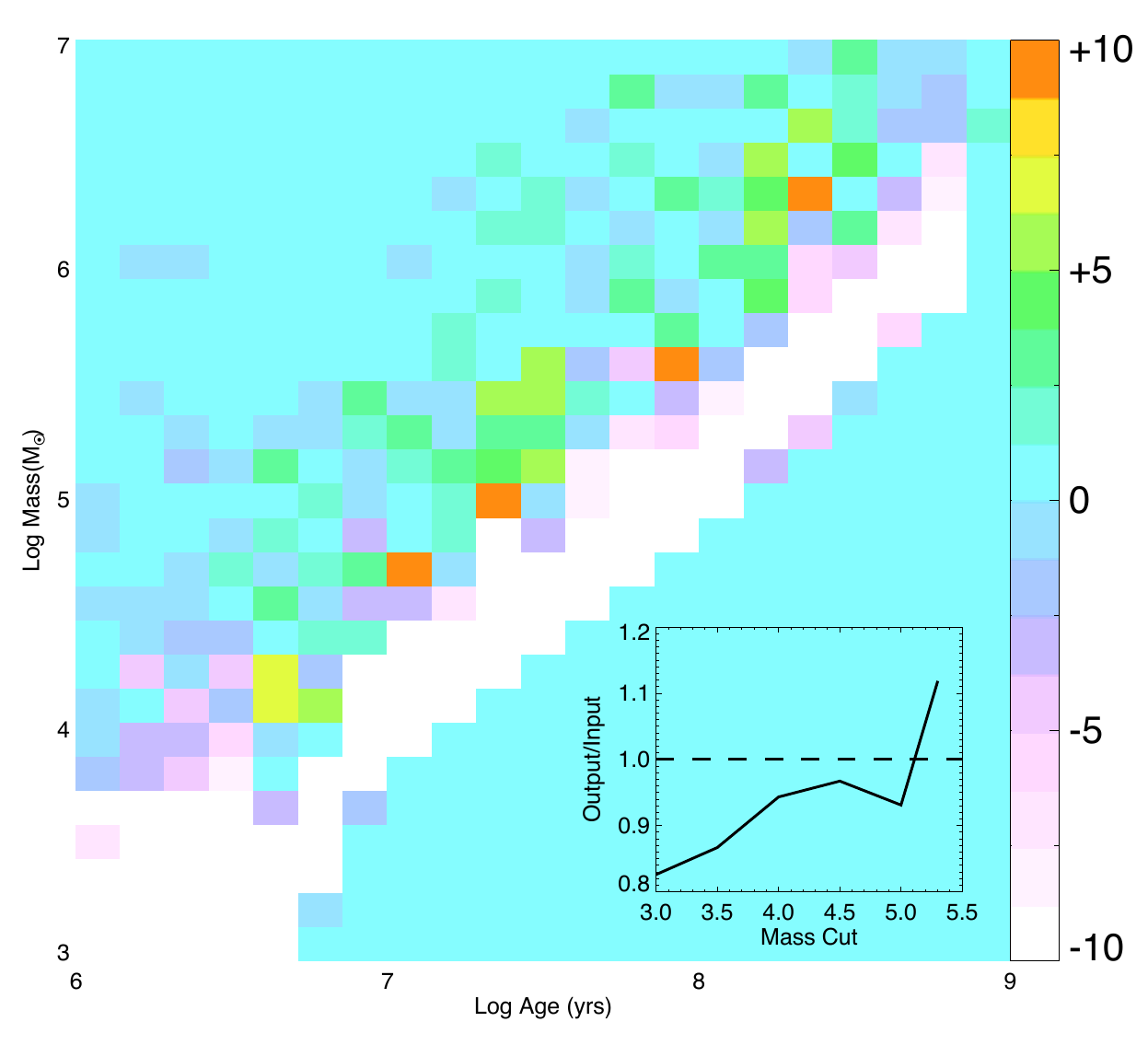}

\caption{Density map showing the difference in cluster parameters between our model cluster population and the results of the cluster fitting procedure. Positive values indicate values where we over estimate the number of clusters. The inset panel shows the mass of the input model population compared to the mass obtained by the cluster fitting procedure for various mass cuts for clusters with ages less than 10Myr.}
   \label{fig:comp}
\end{figure}

We have compared the age-mass distribution of the model cluster population to the results of the SSP fitting routine in Figure \ref{fig:comp}. The colour scale represents the difference in the number of clusters found with a given age and mass compared to the input model population. The white slope across the Figure shows a large deficit of clusters, this is due to the stringent limits we apply to our clusters, accepting only clusters with low errors ($<0.2$). Any faint clusters will have large errors and so we reject the fit. Figure \ref{fig:comp} shows that young clusters are not fitted as we assume we will not detect young embedded clusters. The inset in Figure \ref{fig:comp} shows a comparison of the mass in clusters above a mass limit, comparing the results of cluster fitting to the known values of the input model cluster. For mass cuts below $10^5$\msol\ we tend to underestimate the mass above the mass cut. This implies that measuring the total mass of stars within clusters would be underestimated due to the effects of the fitting procedure. This effect is small however and for the mass cut used in our study of NGC 3256 results in an underestimation by 5\%. For the largest mass cut used we in fact over estimate this mass but only by 10\%, though it should be noted with higher mass cuts we are more likely to encounter errors due to the stochastic sampling of massive young clusters.

The test we have performed here to estimate the error associated with the SSP fitting does not take into account the intrinsic problems with either the Padova or Geneva SSP models. These synthetic stellar evolution tracks do not perfectly represent real star clusters, especially at young ages. Even for models without stellar rotation or binary stars the uncertainties associated with massive stars can be large, and when these effects are included the errors increase. The true affect of fitting to inaccurate SSP models is ultimately hard to quantify and the analysis we have carried out can only be considered as a rough estimate of this effect. 

\subsubsection{Error Due to an Unknown Metallicity}
\label{sec:error_metallicity}

In determining the cluster parameters of age, mass and extinction we have fitted the photometric data to a SSP model of a certain metallicity. Using optical spectroscopy, T07a found that the metallicity of clusters and H{\sc ii} regions in NGC~3256 was between 1 and 1.7~\zsol.  The Padova SSP models used have only two metallicities in/near this range, namely solar and twice solar.  Here we test the effect of our metallicity assumptions on our derived value of $\Gamma$.

\begin{figure}
   \includegraphics[width= 84mm]{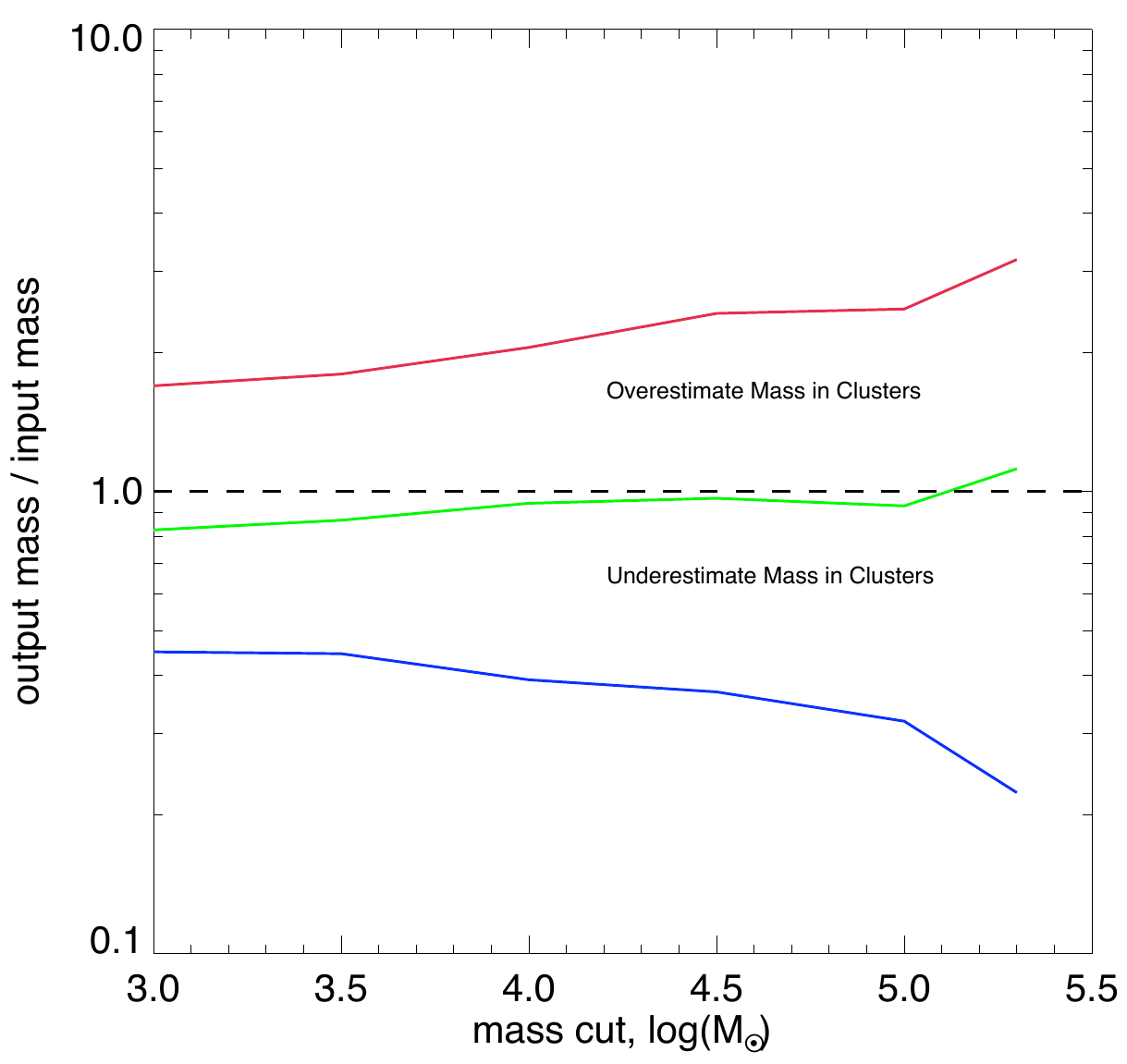}

\caption{The effect of under or over estimating the metallicity of clusters on the measurement of $\Gamma$.  The red line represents the results of fitting a population of solar metallicity with twice solar metallicity SSP models, while the blue line shows the opposite (fitting a twice solar metallicity population with solar models).  The green line is the same as the inset of Fig.~\ref{fig:comp}.}
   \label{fig:metcomp}
\end{figure}

Just as we did in \S~\ref{sec:errs1} we produced model cluster populations based on the solar metallicity Padova tracks, however we then tried to fit these data to the twice solar Padova SSP tracks. We also produced a model population based on the twice solar Padova tracks and fitted them to the solar Padova tracks in an attempt to examine the effects of over or under estimating the metallicity. These results are shown in Figure \ref{fig:metcomp}, the red line shows what happens when we overestimate the metallicity, using a twice solar metallicity track to fit clusters with a solar metallicity. In this case we overestimate the mass in cluster above a particular mass cut. This overestimation changes a little depending on the particular mass cut but is approximately a factor of two for the case of NGC 3256. 


\subsubsection{Error Due to Selection Effects}\label{sec:seleffs}

In measuring $\Gamma$ we need to know how efficient we are at detecting massive, bright clusters, which we use to infer the total mass of stars in clusters. Measuring this efficiency is not trivial and a rigourous, quantitative discussion is not included here. Instead we examined our HST images of NGC 3256, overlaying the objects selected from the B band images. Assuming our detection was perfect all the brightest objects should have been identified. In reality this is not possible as dust will obscure some bright clusters and crowding makes identification of individual clusters difficult. 

\begin{figure}
   \includegraphics[width= 84mm]{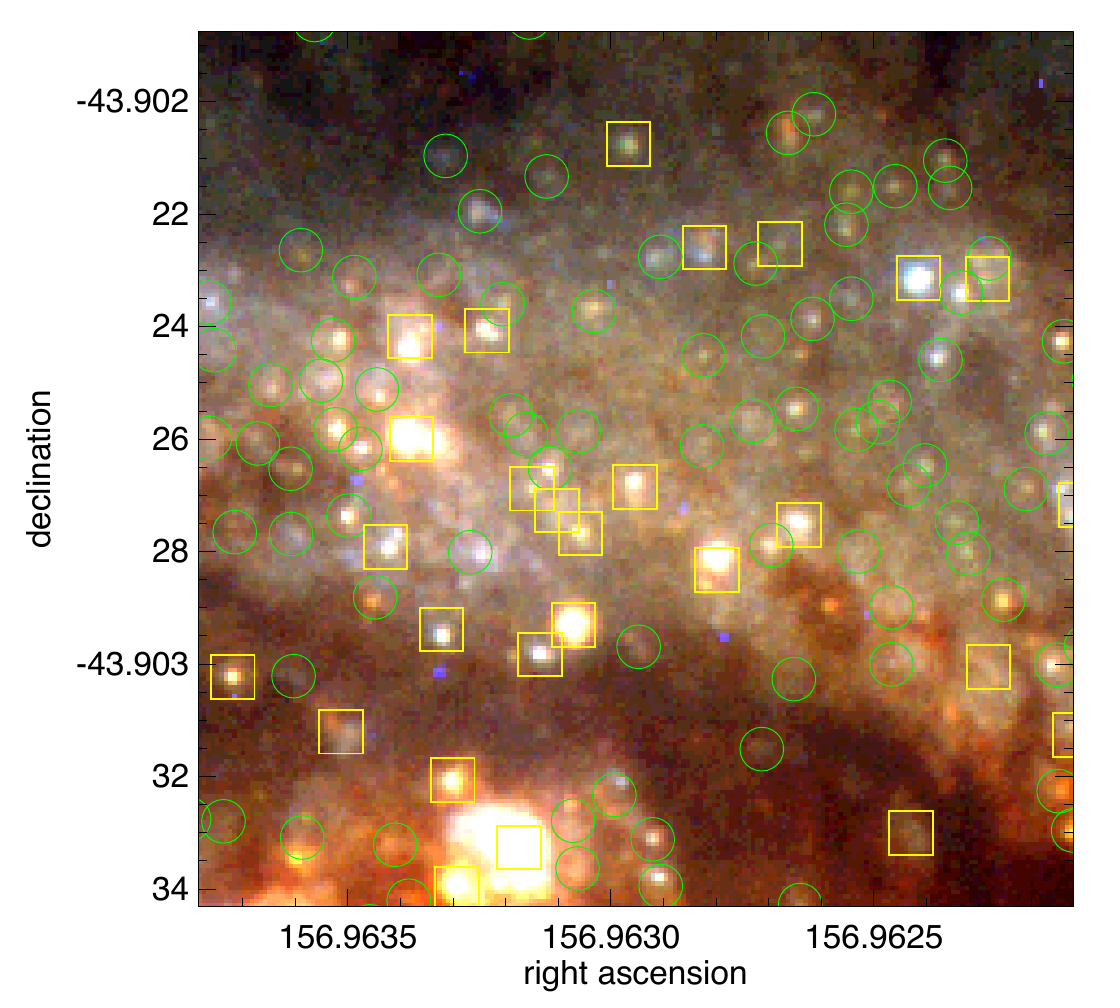}

\caption{Colour composite image of a particularly crowded section of NGC 3256. Objects detected in the B band are circled in green. Objects in yellow boxes passed our criteria for a \emph{good} fit to the SSP models.}
   \label{fig:close}
\end{figure}

We selected the \emph{SExtractor} detection parameters to minimise the chance of two close clusters being identified as a single object. A typical field from a crowded region of NGC 3256 is shown in Figure \ref{fig:close}, and we have overlaid the detected clusters with green circles, those clusters which passed our criteria for a \emph{good} fit have been identified by yellow boxes. The brightest of clusters are all selected and have \emph{good} fits to SSP models. There are several clusters which we detect but do not have \emph{good} fits, on examining these clusters we find that they have either one or two wavebands with errors greater than 0.2 mag and are thus rejected, and labelled as \emph{poor} fits. 

We see from Figure \ref{fig:close} that the actual detection of clusters is not limiting our cluster sample, the ability to accurately fit SSP models is a larger effect. However in our synthetic cluster population models we apply the same criterion to determine \emph{good} fits from \emph{poor} fits. We are very severe in the limits we apply, disregarding any cluster which has an error greater than 0.2 mag in any of the 4 wave bands. It is this criterion which limits our final sample, and is taken into account by modelling cluster populations. We can test this by looking at the fraction of bright, young, massive clusters we reject as \emph{poor} fits in both the real sample and synthetic cluster populations. We examined the percentage of clusters with accepted \emph{good} fits, for clusters younger than 10 Myr and with masses greater than $10^{4.7}$ \msol.  In our sample from NGC 3256 this percentage was 47\% whilst in our synthetic population the result was higher at 76\%. 

The discrepancy in these results could have several causes, firstly our synthetic clusters have a more conservative estimate on the level of dust extinction than we might expect. With a greater level of dust extinction clusters become fainter, photometric errors become larger and the chance we reject the cluster fit increases. Secondly the synthetic cluster population is entirely theoretical and so poor photometry from overcrowding or from misaligned apertures is not taken into account. We can only estimate the percentage of clusters we either miss altogether or discount due to a \emph{poor} cluster fit. An estimate of 50\% is the worst case scenario but some of this is accounted for by our model cluster populations to which we apply the same stringent criteria for \emph{good} cluster fits. Taking this into account we estimate that roughly 20\% of clusters are either not detected or rejected as \emph{poor} cluster fits in addition to the fraction we account for based on the synthetic cluster population models.



\begin{table}
		\caption{A list of the possible sources of uncertainty in measuring $\Gamma$ for NGC 3625. The effect of each source is expressed as the factor applied to correct the measure value of $\Gamma$.}
	\begin{center}
		\begin{tabular}{lcc}
			\hline
			Source & Variation & Effect \\ 
			\hline
			\hline
			CMF slope ($\alpha$) & $\alpha = 1.8 - 2.2$ & $0.66-2.38$ \\ 
			Schechter Function & M$_{*} = 10^{5-7}$\msol & 3.72 - 1.33 \\ 
			SSP Fitting & - & 1.04  \\ 
			Metallicity & Z = 0.5\zsol - 2\zsol & 0.5 - 2 \\ 
			Selection & - & 1.25 \\
			\hline
		\end{tabular}

		\label{tab:sumerrs}
	\end{center}
\end{table}

\subsection{The Value of $\Gamma$ in NGC 3256}

We are in a position to fully calculate the fraction of stars in clusters ($\Gamma$) for NGC 3256, including all the possible souces of error. Initial inspection of our data for NGC 3256 revealed a value of $0.117$. We must account for the over or underestimation introduced when we fit data to cluster models. In \S~\ref{sec:errs1} we calculated that we would underestimate the mass above our mass limit by 5\%, however to make this calculation we made model cluster photometry based on SSP models, and then compared this to the same models. In reality clusters are not perfectly modeled by any SSP and this underestimation is likely larger in reality. We have been rather sceptical in assuming that we only recover $80\pm10\%$ of the mass when we use this cluster fitting method. This increases the value of $\Gamma$ from $0.12\pm0.02$ to $0.15\pm0.03$.

If we assume all our clusters were solar metallicity the we expect to recover 96\% of the mass using a mass cut of $10^{4.7}$\msol.  A more likely scenario is that the cluster population has metallicities in the range $1-2$ Z\ensuremath{_{\small{\sun}}}, this will affect the total mass estimated to be in clusters compared to the actual mass, as shown in Figure \ref{fig:metcomp}. As the metallicity of each individual cluster is unknown we cannot be certain if the mass in clusters calculated is an under or over estimation of the true mass. We expect clusters in NGC 3256 to be closer to a solar metallicity than twice solar, and so estimate that our cluster fitting method underestimates the total mass by a factor of  $0.8\pm^{0.2}_{0.3}$ taken from figure \ref{fig:metcomp} for the mass cut used in our study. This further increases $\Gamma$ from $0.15\pm0.03$ to $0.18\pm^{0.06}_{0.08}$.

We also need to consider the fraction of clusters that were not detected or thrown out as \emph{poor} cluster fits. We estimated this fraction to be 20\% in section \ref{sec:seleffs}. Combining these effects we arrive at a more assured value for $\Gamma$ than the value the data initially suggests. The value used throughout the remainder of this study is $0.23\pm^{0.07}_{0.10}$.

\section{Results from Other Data-sets}
\label{sec:results}

Fitting photometric observations to cluster evolutionary tracks is very time consuming. This study does not attempt to replicate the workings carried out on the clusters of NGC 3256 on numerous other galaxies. Instead we have used previous results from other studies, in an effort to calculate the ratio of stars forming within clusters. The various data sources are discussed below.



\subsection{NGC 1569}

We use the data set of \citet{anders04b}, which includes ages and masses for 161 clusters in the galaxy NGC 1569. Just as for NGC 3256 we consider clusters younger than 10 Myrs. With the small distance to NGC 1569 \citet{anders04b} was able to observe many low mass clusters, and we can use a low mass cut to estimate the total mass in clusters. In order to avoid measurements which are affected heavily by stochastic sampling of the ICMF we must use the lowest mass limit possible whilst being wary of the observational limits. We chose three lower mass limits, $10^{3.0}$ \msol, $10^{3.2}$ \msol\ and $10^{3.4}$ \msol. In the same manner as we did for NGC 3256 we extrapolated the mass found above this limit up to the total mass in clusters, this resulted in consistent answers from all three of these limits. Averageing these three results we calculated the total mass in clusters to be $3.52 \pm 0.05 \times 10^{5}$ \msol. Assuming we are unable to observe any embedded clusters younger than three million years the resulting CFR is $0.05$ \msol \ yr$^{-1}$. The star formation rate is taken from an \ha measurement from \citet{moustakas06} and the \ha SFR calibration in \citet{kennicutt98}, using the same adopted distance to NGC 1569 as \citet{anders04b} of 2.2Mpc. Giving a SFR of $0.36 \pm 0.02$ \msol\ yr$^{-1}$. $\Gamma$ is simply the ratio of the CFR to the SFR, $13.9\pm0.8\%$.



\subsection{NGC 6946}

We used data from \citet{larsen02} who quoted ages and masses for 90 clusters in NGC 6946. Considering clusters younger than 10 Myrs and a lower mass limit of $10^{3.5}$ \msol \ we calculated the mass in clusters above this limit, giving $1.284 \times 10^5$ \msol. The fraction of mass expected above this mass limit is calculated as in previous sections and results in a total inferred cluster mass of $1.95 \pm 0.23 \times 10^5$ \msol. We continue to assume that clusters younger than 3 Myrs are embedded an unobservable and so have an age range of 7 Myrs, thus giving a CFR of $0.022 \pm 0.003$ \msol \ yr$^{-1}$. 


The data of  \citet{larsen02} only comes from one pointing of the HST WFPC2 camera and does not include the whole galaxy. To calculate $\Gamma$ we need the SFR across this region only. This was found using the area covered by the WFPC2 and the SFR density taken from \citet{larsen02}, giving a SFR of 0.1725  \msol yr$^{-1}$. Consequently $\Gamma$ is $12.5\pm^{1.8}_{2.5} \%$, although we note this HST pointing included the centre of NGC 6946 and so the SFR might be higher that the global average we used.

\subsection{Small Magellanic Cloud}

We have used the data set generated by \citet{hunter03}, a catalogue of ages and masses for 191 clusters in the SMC. This catalogue is thought to be incomplete for clusters younger than 10 Myrs, so we calculated the mass in clusters between 10 and 100 Myrs. With the SMC lying at a distance of $61 \pm 3$ kpc \citep{hilditch05} the data set is complete to very low cluster masses. 

Using lower mass limits of $10^{2.6}, 10^{2.8}, 10^{3.0}\ \&\ 10^{3.2}$ \msol\ to calculate the mass in total mass in clusters gave consistent results of $1.59 \pm 0.03 \times 10^5$ \msol. We assumed a constant CFR and SFR over the age range 10-100 Myrs, giving a CFR of $1.77 \pm 0.03 \times 10^{-3}$ \msol\ yr$^{-1}$. The SFR which we assume to be constant over the last 100 million years is taken from the extinction corrected \ha luminosity value of \citet{kennicutt86}, converted to a SFR of $0.043$ \msol yr$^{-1}$ using the Equations given by \citet{kennicutt98}. The resulting value for $\Gamma$ is $4.2\pm^{0.2}_{0.3}\%$. 

Our value for $\Gamma$ agrees extremely well with the value calculated by \citet{gieles08} of 3-5\%, derived using size-of-sample effects on the same cluster sample. \citet{kruijssen08} looked at the ratio of clustered to field stars in the SMC and quote a minimum $\Gamma$ of $0.5\%$ and suggest a more reasonable value of $10\%$ as their best result.

\subsection{Large Magellanic Cloud}

Ages and masses for 748 clusters in the LMC were taken from the data set of \citet{hunter03}. Just as in the case of the SMC this catalogue is believed to be incomplete for clusters younger than 10 millions years and so we studied clusters in the age range $10-100$ million years. 

With the LMC being a relatively nearby galaxy at a distance of only 58.5 kpc \citep{macri06} the sample is complete to low mass and we calculated the mass of clusters above four mass limits $10^{2.6}, 10^{2.8}, 10^{3.0}\ \&\ 10^{3.2}$ \msol. The fraction of mass expected above these limits was found as described in previous sections, resulting in a total inferred mass in clusters of $6.31 \pm 0.20 \times 10^5$ \msol. We assume a constant SFR and CFR over the 90 million year period considered, resulting in a calculted CFR of $7.01 \pm 0.22 \times 10^{-3}$ \msol\ yr$^{-1}$. 

The SFR ($0.12$ \msol\ yr$^{-1}$) and the galaxy area (79 kpc$^2$) are both taken from \citet{larsen00}. $\Gamma$ for the LMC was thus found to be $5.8\pm0.5\%$.


\subsection{The Milky Way (The Solar Neighbourhood)}

\citet{lada03} estimate the rate of embedded cluster formation within 2.0kpc of the sun to be between $2-4$ Myrs$^{-1}$ kpc$^{-2}$, and claim the average mass of an embedded cluster to be 500\msol. \citet{lada03} go on to estimate the number of embedded clusters that survive, with approximately 7\% surviving to the age of the Pleiades. 

We convert the rate of embedded cluster formation to a mass formation rate assuming the average cluster mass of 500\msol, giving $0.15$\msol\ yr$^{-1}$ in the solar neighbourhood. We calculate a value for $\Gamma$ by assuming the SFR is equal to the rate of embedded cluster formation. Comparing this to the mass found in older clusters, 7\% though estimates of this percentage varies between 4\% and 14\% depending on the numbers used.  \citet{roberts57}, \citet{millerscalo78}, and \citet{adamsmyers01} also derived a similar percentage of star-formation in clusters for the solar neighbourhood.


\subsection{M83}

\citet{harris01} presented a catalogue of 45 massive star clusters in the centre of M83, with reliable ages, masses and extinction values. Using this catalogue and a lower mass limit of $10^{2.8}$ \msol\ we calculated the total mass in clusters younger than 10 million years to be $7.24 \pm 0.23 \times 10^5$ \msol. To calculate the CFR we assume that clusters younger than 3 million years old are embedded and so are unobservable. The resulting CFR is $0.10 \pm 0.01$ \msol\ yr$^{-1}$.

The study of \citet{harris01} only covers the central region of M83 and so we used an \ha image of M83 to measure the \ha from the same region studied by \citet{harris01}. We used archival narrowband and R-band images of M83 taken as part of the SINGS survey \citep{meurer06}. We measured the background subtracted flux for an identical area as that studied by \citet{harris01} and converted this flux to a SFR using the calibrations of \citet{kennicutt98}, resulting in a SFR of $0.23 \pm 0.03$ \msol\ yr$^{-1}$.  We correct this value for extinction using the reddening curves of \citet{calzetti01b}. The extinction value was taken as the averaged extinction of all clusters in the \citet{harris01} sample, giving a correction of 1.71, the resulting dust corrected SFR is $0.39 \pm 0.06$ \msol\ yr$^{-1}$. 

With the CFR and SFR known we calculate the value of $\Gamma$ to be $26.7\pm^{5.3}_{4.0}\%$.
 
\subsection{The Antennae Galaxies}

Ages and masses have been published for 752 clusters in the Antennae Galaxies by \citet{anders07}. We attempted to apply the same calculations as we did for other cluster populations, as described above. However using different lower mass limits we obtained a range of values for $\Gamma$, from $60\%$ to $>100\%$; assuming a star formation rate of 20~\msol yr$^{-1}$ \citep{zhang01}. Obviously values in-excess of $100\%$ are unphysical and this problem is exasperated  by the range of quoted SFR's for the Antennae Galaxies which can range from $5-20$ \msol yr$^{-1}$ \citep{zhang01,knierman03}.

We investigated the slope of the mass function for young clusters and found that this data set presents a very shallow slope of $\alpha\simeq1.6$. Given that mass function slopes are usually in the range $1.8<\alpha<2.2$ \citep{degrijs03c} a slope this shallow represents a very unusual population, unfortunately \citet{anders07} do not calculate a similar mass function for these clusters. Correcting for this shallow mass function gave results for $\Gamma$ that were still above $50\%$ and did vary significantly with the assumed lower mass limit. In comparison we noted that \citet{fall05} conclude that at least $20\%$ but possible all clusters form in clusters; the value for $\Gamma$ may well be high for the Antennae.

Given the range of values we can calculate for $\Gamma$ based on the \citet{anders07} data set and the extremely shallow mass function of young clusters in this sample combined with the range in quoted SFR's for the Antennae we do not publish a value for this system. It serves as a reminder of how difficult these calculations can be and how dependent they are on accurate knowledge of the SFR.  


\section{The Variation of $\Gamma$ with SFR}
\label{sec:implications}

\begin{figure}
   \includegraphics[width= 84mm]{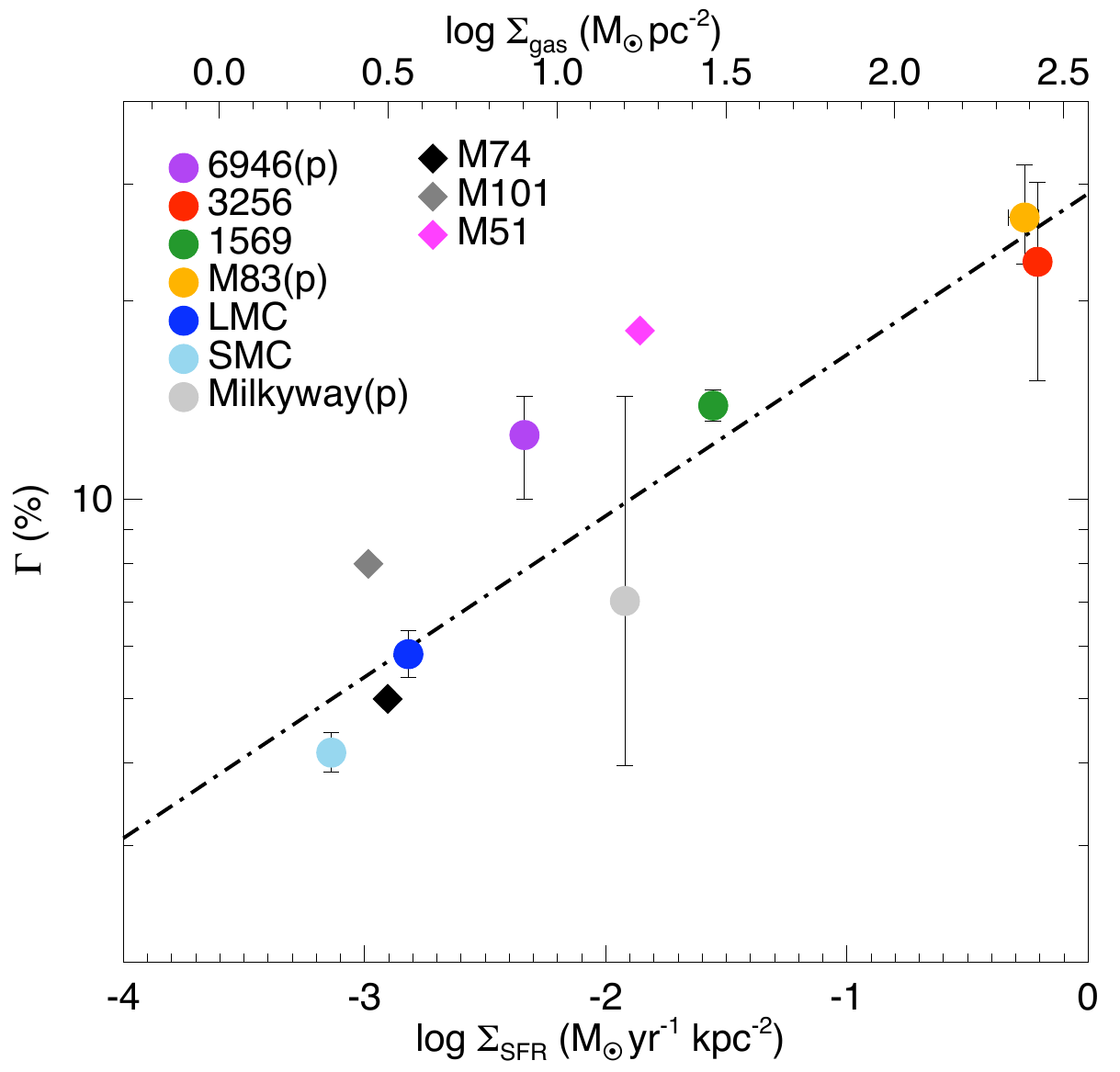}
\caption{Graph showing $\Gamma$ against star formation rate density for all galaxies examined in our study. The dot-dashed line shows a least-squares best fit power law to the distribution. Diamonds show data points taken from \citet{gieles09b} and are not included in calculating the best fit to the data.}
   \label{fig:final2}
\end{figure}

In Table \ref{tab:res} we present all the information used hereafter, including the values of $\Gamma$, star formation rate (SFR) and area over which these quantities were measured. The superscript \emph{p} after the galaxy names indicates that only a partial area of the galaxy was used to derive these results, as described in the previous sections. At this point it is worth noting that because results have been obtained from various data sets the methodology used differs between results. Overall the consistency of results is thought to be robust as measurements of $\Gamma$ are based on the brightest and easiest clusters to detect which should give consistent results. It is harder to estimate the errors associated with $\Gamma$ based on other data sets. Results shown in Figure \ref{fig:final2} for galaxies other than NGC 3256 and the Milky Way only include errors associated with the uncertainty in the fraction of mass expected above whatever lower mass limit was used. Errors due to uncertainties in the metallicity and SSP fitting have not been estimated as this is difficult without reprocessing the entire data set.

\begin{table}
\caption{Summary of results included in this paper.}
\label{tab:res}
\begin{tabular}{@{}l c c c}
\hline
\hline
Galaxy & SFR & A & $\Gamma$ \\
 & (\msol \ yr$^{-1}$) & (kpc$^{2})$ & (\%) \\
\hline
NGC 1569 		& 0.3626	& 13		& $13.9\pm0.8$	\\
NGC 3256 		& 46.17	& 74.85	& $22.9\pm^{7.3}_{9.8}$	\\
NGC 5236 (M83)$^{p}$& 	0.3867& 0.7077	& $26.7\pm^{5.3}_{4.0}$	\\
NGC 6946$^{p}$	& 0.1725	& 37.49	& $12.5\pm^{1.8}_{2.5}$	\\
LMC				& 0.1201	& 79		& $5.8\pm0.5$	\\
SMC				& 0.0426	& 58.55	& $4.2\pm^{0.2}_{0.3}$	\\
Milky Way$^{p}$	& 0.1508	& 12.56	& $7.0\pm^{7.0}_{3.0}$	\\
\hline
\end{tabular}
\end{table}

Figure \ref{fig:final2} shows the relation between the SFR density ($\Sigma_{SFR}$) and $\Gamma$ for all galaxies discussed in this paper. In addition we plot results for 3 galaxies from \citet{gieles09b} in which CFRs (and subsequently $\Gamma$) are calculated via comparison with empirical luminosity functions. It is immediately obvious that $\Gamma$ increases with the SFR density. Over three orders of magnitude a power-law relationship holds and the dot-dashed line shows a least-squares fit to the data of the form, $\Gamma \propto \Sigma_{SFR}^{\alpha}$. The numerical version of the derived relationship is displayed in Equation \ref{eqn:pl}.

\begin{equation}
\label{eqn:pl}
\Gamma(\%) = (29.0\pm{6.0}) \Sigma_{SFR}^{0.24\pm0.04} (\msol \ yr^{-1}\ kpc^{-2})
\end{equation}

From this Equation we can predict at what density we might expect to see all stars in clusters, effectively a $\Gamma$ of 100\%, this would occur at a SFR density of $7.5\times10^3$ \msol\ yr$^{-1}$ kpc$^{-2}$, a rather high value. Interestingly we can integrate this value over the time it takes a cluster to form, assuming this takes roughly three million years we expect $2.25 \times 10^{4}$ \msol\ pc$^{-2}$, which is approximately the density of a typical cluster. 

Although the results taken from \citet{gieles09b} are not calculated in the same manner as we have for other data sets these points fit the trend quite well, particularly at low SFR densities. It should be noted that our calculation of the best fit power law does not include these additional data points, given the different method used to determine these results, they are shown here to demonstrate the consistency of our results. 

We have investigated whether the observed correlation (Eqn.~\ref{eqn:pl}) could be due to a systematic effect, namely if the mass function index varied with environment.  Assuming the LMC cluster distribution is well approximated with $\alpha=2$, an index of $\alpha \sim 1.2$ would be required in order to bring the results of NGC~3256 into agreement with the LMC.  This kind of gross deviation is ruled out by our direct observations of the cluster population (e.g. Fig.~\ref{fig:massfn}).

A similar conclusion (that more stars are formed in bound clusters with increasing SFR density) has been reached based on the fraction of light observed in clusters relative to the host galaxy.  Specifically, Meurer et al.~(1995) and Zepf et al.~(1999) found this for merging/starburst galaxies and \citet{larsen00} and \citet{larsen04} found this for a larger sample that includes starburst and quiescent galaxies.  While measuring the fraction of light in clusters is significantly easier than deriving $\Gamma$, hence suitable for construction of large surveys, it is affected by the star formation history of the galaxy and possibly by differential extinction effects.  The exact relation between $\Gamma$ and the fraction of light observed in clusters will be modelled in a future paper.



\citet{kennicutt98b} investigated a global Schmidt-Kennicutt law across galaxies, correlating the SFR density to the surface gas density. As $\Gamma$ appears to follow a power-law relationship with the SFR density using the Schmidt-Kennicutt law defined in \citet{kennicutt98b} we can construct a relation between $\Gamma$ and the surface gas density, shown in Equation \ref{eqn:gas}.

\begin{equation}
\label{eqn:gas}
\Gamma(\%) = (4.1\pm1.9) \Sigma_{gas}^{0.34\pm0.07} (\msol \ pc^{-2})
\end{equation}

We show the derived surface gas density corresponding to a given SFR density on the top axis of Figure \ref{fig:final2}. Although our data only charts the value of $\Gamma$ for gas densities in the range $1-300$ \msol\ pc$^{-2}$ it is insightfull to interpret this relation to the more extreme star forming environments. Arp220 represents one of the most active star forming galaxies found \citep{wilson06}, with a global SFR of 240 \msol\ yr$^{-1}$ based on the FIR luminosity of \citet{sanders03} and the calibration of \citet{kennicutt98}. Assuming that Arp220 occupies an area of roughly 1 kpc$^{3}$ \citep{ananth00}, and thus has a surface area of 1 kpc$^{2}$ on the sky we can derive the SFR density is approximately 240 \msol\ yr$^{-1}$ kpc$^{-2}$. Such a high SFR density translates to a value of $\Gamma$ of roughly 85\% using Equation \ref{eqn:pl}, indicating that almost all the star clusters in Arp220 would be expected to survive the embedded phase, and the CMF would be almost identical to the true underlying initial CMF. Hence, there are regions where we would expect most/all young stars to be found in clusters.

\section{Conclusions}
\label{sec:conclusions}

Over the course of this paper we have shown that it is possible to accurately measure the fraction of stars found within young clusters, a parameter we have termed $\Gamma$. This is achieved by obtaining ages and masses for the cluster population of NGC 3256 through multiple waveband photometry and comparison with synthetic stellar population models. We have also been able to calculate $\Gamma$ for several other galaxies using published cluster populations. We examined how $\Gamma$ varies with the star formation rate of the host galaxy and we summarise our conclusions below.

\begin{enumerate}
\item The cluster formation history of NGC 3256 shows an increase in the star formation rate over the last 100 million years, most likely caused by the ongoing merger of the two progenitor spiral galaxies. The CMF for NGC 3256 is best described by a power law with slope $\alpha = 1.85\pm0.12$.

\item $\Gamma$ may be calculated directly if an accurate cluster population is known, however there are several possible sources of error in making these calculations. The effect of SSP fitting must be taken into account and an unknown metallicity may produce additional uncertainties. Selection effects may also alter the value of $\Gamma$ but this can be estimated. The calculated $\Gamma$ does depend on the form of the cluster mass function used, we assume a simple power law function with slope $\alpha = 2.0$. We have shown that a Schechter function produces similar but higher values, however this effect is small if the \mstar is high (\mstar $> 10^6$\msol).

\item We found a weak positive correlation of $\Gamma$ with the total star formation rate. We find a strong relation between $\Gamma$ and the star formation rate density, which we have written as a power law type relation, $\Gamma(\%) = (25.5\pm{6.0}) \Sigma_{SFR}^{0.22\pm0.05}$ (\msol \ yr$^{-1}$\ kpc$^{-2}$). This is similar to that found by \citet{larsen00} for the fraction of U-band light in clusters relative to the total galaxy.  This result can also be interpreted as a correlation with the surface gas density through the Schmidt-Kennicutt Law \citep{kennicutt98b}. This implies that either clusters born in high density environments are more resistant to disruption in the embedded phase, or the environment changes the fraction of stars born in clusters.

\end{enumerate}

Measuring $\Gamma$ is not trivial and it does have several sources of error which must be taken into account. However it is possible to make these calculations which may be further refined with consistent measurements using the same instruments, detection methods and cluster fitting models. This study is intended to show that if all sources of uncertainty are taken into account and measured then calculating $\Gamma$ is possible.  Future work, including larger homogeneous samples and deeper observations to constrain the form of the cluster mass function to lower masses, will be needed to conclusively investigate the intriguing relation between $\Gamma$ and the SFR density of the host galaxy.

\section*{Acknowledgments}

We thank S\o ren Larsen and Mark Gieles for valuable discussions and suggestions.  NB is supported by an STFC Advanced Fellowship.


\onecolumn
\begin{landscape}
\begin{longtable}{l c c c c c c c c c c c c c c c c c c c c}
\caption[hoops]{Cluster photometry and parameters for the 276 \emph{good} cluster fits.} \label{tab:fulldat} \\

 \hline\hline     & $\Delta$ra & $\Delta$dec & \multicolumn{8}{c}{Photometry} & \multicolumn{3}{c}{E(B-V)} & \multicolumn{3}{c}{log age (yrs)} & \multicolumn{3}{c}{log mass (\msol)} \\ 
ID & (s)&  (arcsec) &  U & $\Delta$U & B & $\Delta$B & V & $\Delta$V & I & $\Delta$I & min & best & max & min & best & max & min & best & max \\ 
(1) & (2) & (3) & (4) & (5) & (6) & (7) & (8) & (9) & (10) & (11) & (12) & (13) & (14) & (15) & (16) & (17) & (18) & (19) & (20) & \\ \hline
\endfirsthead 
\hline\hline 
 & $\Delta$ra & $\Delta$dec & \multicolumn{8}{c}{Photometry} & \multicolumn{3}{c}{E(B-V)} & \multicolumn{3}{c}{log age (yrs)} & \multicolumn{3}{c}{log mass (\msol)} \\
ID & (s) &  (arcsec) &  U & $\Delta$U & B & $\Delta$B & V & $\Delta$V & I & $\Delta$I & min & best & max & min & best & max & min & best & max \\
(1) & (2) & (3) & (4) & (5) & (6) & (7) & (8) & (9) & (10) & (11) & (12) & (13) & (14) & (15) & (16) & (17) & (18) & (19) & (20) & \\ \hline
\endhead
\hline
\endfoot
     1 & 51.61 & 28.11 & 22.79 &  0.18 & 23.93 &  0.09 & 23.14 &  0.07 & 22.40 &  0.07 &  0.18 &  0.42 &  0.70 &  6.92 &  6.79 &  6.92 &  4.12 &  4.24 &  4.37 \\
      2 & 50.25 & 24.82 & 20.82 &  0.06 & 21.98 &  0.10 & 21.51 &  0.05 & 21.02 &  0.07 &  0.00 &  0.04 &  0.22 &  6.94 &  6.85 &  6.94 &  4.47 &  4.53 &  4.64 \\
      3 & 52.02 & 27.63 & 20.25 &  0.03 & 21.07 &  0.01 & 20.81 &  0.01 & 20.17 &  0.01 &  0.00 &  0.02 &  0.14 &  6.98 &  6.90 &  6.98 &  4.87 &  4.90 &  5.05 \\
      4 & 51.20 & 26.78 & 22.58 &  0.13 & 23.30 &  0.05 & 22.60 &  0.06 & 21.80 &  0.05 &  0.26 &  0.50 &  0.70 &  6.90 &  6.77 &  6.90 &  4.43 &  4.54 &  4.62 \\
      5 & 51.81 & 27.61 & 22.00 &  0.09 & 22.61 &  0.02 & 22.28 &  0.04 & 21.48 &  0.03 &  0.12 &  0.22 &  0.26 &  7.75 &  7.56 &  7.75 &  4.44 &  5.31 &  5.44 \\
      6 & 50.23 & 25.06 & 21.12 &  0.06 & 22.19 &  0.08 & 21.62 &  0.05 & 21.20 &  0.06 &  0.08 &  0.24 &  0.44 &  6.85 &  6.75 &  6.85 &  4.49 &  4.56 &  4.65 \\
\end{longtable}
(1) Gives the Object ID number referred to through out the paper. (2) and (3) give the right accession and declination offsets form R.A. = $10^{h}20^{m}00^{}s$; decl. = $-43 54 00$ (J2000.0). Cols. (4-11) give the U, B, V \& I band photometry with associated errors. (12-14) gives the dust extinction of the object in terms of E(B-V) after correcting for foreground galactic extinction as derived from our cluster fitting method.  We show the most likely value as well as the maximum and minimum acceptable results. (15-17) shows the cluster age in log (yrs), once again we show the minimum, best and maximum values from the cluster fitting. (18-20) log cluster mass with maximum, minimum and best values from the cluster fitting. A full version of this table is available with the online edition of this article.
 \end{landscape}
 \twocolumn

\bsp

\label{lastpage}

\end{document}